\documentclass[nolineno]{jfm}

\usepackage{graphicx}
\usepackage{newtxtext}
\usepackage{newtxmath}
\usepackage{natbib}

\usepackage{graphicx}
\usepackage{amsmath,amssymb}

\usepackage{float}
\usepackage{hyperref}
\hypersetup{
    colorlinks = true,
    urlcolor   = blue,
    citecolor  = black,
}

\newcommand{\rayleigh}{Ra}
\newcommand{\prandtl}{\sigma}
\newcommand{\reynolds}{Re}
\newcommand{\rossby}{Ro}
\newcommand{\ekman}{Ek}
\newcommand{\taylor}{Ta}
\newcommand{\ez}{\mathbf{\widehat{e}}_z }
\newcommand{\ex}{\mathbf{\widehat{e}}_x }
\newcommand{\ey}{\mathbf{\widehat{e}}_y }
\usepackage[percent]{overpic}
\usepackage[symbol]{footmisc}

\title{Bridging the Rossby number gap in rapidly rotating thermal convection}

\author{Adrian van Kan\aff{1}
  \corresp{\email{adrianvankan@gmx.de}},
 Keith Julien\aff{2}\footnote{Deceased on 14th April 2024.}, Benjamin Miquel\aff{3}, Edgar Knobloch\aff{1}
 }

\affiliation{\aff{1}{Department of Physics}, University of California at Berkeley, Berkeley, CA 94720, USA.
\aff{2}Department of Applied Mathematics, University of Colorado, Boulder, CO 80309, USA.
\aff{3} CNRS, \'Ecole Centrale de Lyon, INSA Lyon, Universit\'e Claude Bernard Lyon 1, Laboratoire de M\'ecanique des Fluides et d’Acoustique, UMR5509, F-69134 \'Ecully, France.
}

\begin{document}
\maketitle

\begin{abstract}
Geophysical and astrophysical fluid flows are typically driven by buoyancy and strongly 
constrained at large scales by planetary rotation. Rapidly rotating Rayleigh--B\'enard 
convection (RRRBC) provides a paradigm for experiments and direct numerical simulations 
(DNS) of such flows, but the accessible parameter space remains restricted to moderately fast 
rotation rates (Ekman numbers $\ekman \gtrsim 10^{-8}$), while realistic $\ekman$ for 
geo- and astrophysical applications are orders of magnitude smaller. On the other hand, previously 
derived reduced equations of motion describing the leading-order behaviour in the limit of very 
rapid rotation ($\ekman\to 0$) cannot capture finite rotation effects, and the physically most 
relevant part of parameter space with small but finite $\ekman$ has remained elusive. Here, 
we employ the rescaled rapidly rotating incompressible Navier--Stokes equations 
(RRRiNSE) -- a reformulation of the Navier--Stokes--Boussinesq equations informed by the 
scalings valid for $\ekman\to 0$, recently introduced by \citet{julien2024rescaled} -- to 
provide full DNS of RRRBC at unprecedented rotation strengths down to $Ek=10^{-15}$ 
and below, revealing the disappearance of cyclone--anticyclone asymmetry at previously 
unattainable Ekman numbers ($\ekman\approx 10^{-9}$). We also identify an overshoot in 
the heat transport as $\ekman$ is varied at fixed $\widetilde{Ra}\equiv Ra\ekman^{4/3}$, 
where $Ra$ is the Rayleigh number,
associated with dissipation due to ageostrophic motions in the boundary layers. 
The simulations validate theoretical predictions based on thermal boundary layer theory 
for RRRBC and show that the solutions of RRRiNSE agree with the reduced equations at very 
small $\ekman$. These results represent a first foray into the vast, largely unexplored parameter 
space of very rapidly rotating convection rendered accessible by RRRiNSE.
\end{abstract}



\section{Introduction}
\label{sec:headings}
 The universe abounds with examples of turbulent flows that are driven by buoyancy and constrained by rotation. These highly complex flows shape our environment across scales: in protoplanetary disks, they mediate the formation of planetesimals \citep{lesur2010angular}; in planetary atmospheres, including on Earth, they carry heat, momentum and moisture, crucially impacting the climate \citep{emanuel1994atmospheric,dauxois2021confronting,siegelman2022moist}; in the Earth's oceans 
 \citep{marshall1999open,cheon2019open}, they transport heat, carbon dioxide, and nutrients, thereby crucially influencing biotopes \citep{severin2014impact}; in planetary and stellar cores, they are responsible for the generation of large-scale magnetic fields via the dynamo instability
\citep{king2010convective,jones2011planetary,schaeffer2017turbulent}; in the subsurface oceans of the gaseous giants' moons, they shape the icy crusts of Europa, Enceladus, Titan, etc. \citep{pappalardo1998geological,mitri2008thermal,nimmoJGRP16,soderlund2024physical}; in stars like the Sun \citep{miesch2000coupling,fan2021magnetic} they are responsible for large-scale magnetic fields as well as rich dynamics such as the 22-year cycle in solar activity.

Such geo- and astrophysical flows, in spite of their differences, typically share two important properties. 
Owing to their turbulent nature, their dynamics results from a broad range of interacting spatial and 
temporal scales. Accordingly, advection by the flow dominates over viscous forces when the Reynolds number 
is large, $\reynolds_H=UH/\nu\gg 1$, based here on the fluid layer depth $H$, a typical velocity of the 
flow $U$, and kinematic viscosity $\nu$. Simultaneously, advection by the flow occurs on time scales much 
longer than the planetary rotation period, as measured by small Rossby numbers 
$\rossby_H=U/(2\Omega H)\ll 1$, where $\Omega$ is the planetary rotation rate. 

Also of particular importance is the ratio of the viscous and Coriolis forces providing an 
\textit{a priori} external parameter referred to as the Ekman number 
$\ekman = Ro_H/Re_H = \nu/(2\Omega H^2)$, 
as well as the Prandtl number $\sigma  = \nu/\kappa$ characterising the fluid under consideration, 
where $\kappa$ is the thermal diffusivity. The strength of the thermal forcing is controlled by the
Rayleigh number $Ra$, which is proportional to the temperature drop $\Delta_T $ across the fluid layer. 
This number can be used to define an alternate \textit{a priori} parameter, the convective Rossby 
number $Ro_{conv}= \sqrt{Ra/\sigma}\, Ek$, 
measuring the strength of the rotation relative to thermal forcing.

Celestial bodies typically feature extreme values for the non-dimensional parameters defined above, 
as summarised in table~\ref{Table:Paramo}: rapid planetary rotation is reflected in very low Ekman 
and Rossby numbers, whereas large Reynolds numbers indicate the highly turbulent character of the flow. 
Such rapidly rotating, highly turbulent flows are challenging to analyze and specifically to simulate 
numerically owing to the sheer number of spatial and temporal degrees of freedom which they involve. 
One common approach for obtaining theoretical predictions at geophysically relevant parameters 
(specifically, $\ekman \rightarrow 0)$ relies upon identifying transitions between different flow 
regimes in a region of more moderate parameters ($\ekman \gtrsim 10^{-8}$) that is amenable to direct 
numerical simulations (DNS) or laboratory experiments, together with the scaling laws for transport 
coefficients valid within each regime. These results are then extrapolated to extreme parameters on 
the assumption that the regimes and scaling laws observed for accessible parameter values extend to 
geo-/astrophysically relevant parameter values. Assuming that no nontrivial transition occurs outside 
the observed moderate parameter interval is a particularly strong assumption, akin to a leap of faith 
due to the current paucity of data at the most extreme parameter values.

\begin{table}
\begin{center}
\begin{tabular}{|l|c|r|c|c|}
\hline
Celestial body & $\ekman$ & $\sigma$  & $\rossby_H$ & $\reynolds_H$ \\
\hline
Earth's outer core & $10^{-15}$ & $0.1$ &  $10^{-7}$ & $10^8$ \\ 
Mercury (core) 
&  $10^{-12}$ & $0.1$  & $10^{-4}$ & $10^8$ \\ 
Jupiter (core)
& $10^{-19}$ & $0.1$  & $10^{-10}$ & $10^9$ \\ 
\ \ \ Europa (ocean)& $10^{-12}$ & $11.0$ & $10^{-2.5}$--$10^{-1.5}$ & $10^{9.5}$--$10^{10.5}$ \\ 
\ \ \ Ganymede (ocean)& $10^{-10}$--$10^{-13}$ & $10.0$  & $10^{-3.5}$--$10^{1.5}$ & $10^{9.5}$--$10^{11.5}$ \\ 
Saturn (core) 
& $10^{-18}$ & 0.1  & $10^{-9}$ & $10^9$ \\
\ \ \  Enceladus (ocean) & $10^{-10}$--$10^{-11}$ & $13.0$  & $10^{-3.5}$--$10^{-1}$ & $10^{7.5}$--$10^{9}$ \\ 
\ \ \ Titan (ocean)
& $10^{-11}$--$10^{-12}$ & $10.0$  & $10^{-3}$--$1$ & $10^{9}$--$10^{11}$ \\ 
Neptune (core) & $10^{-16}$ & 10.0  & $10^{-6}$ & $10^{10}$ \\
Uranus (core) & $10^{-16}$ & 10.0  & $10^{-6}$ & $10^{10}$ \\
Sun (convection zone) & $10^{-15}$ & $10^{-6}$ & $10^{-3}$ & $10^{12}$\\
\hline
\end{tabular}
\end{center}
\caption{\raggedright
Non-dimensional parameter estimates for planetary \citep{schubert2011}, satellite  \citep{soderlund2019ocean} and stellar interiors \citep{miesch2000coupling,garaud2008dynamics}. 
Estimates of the Rossby number are derived from the relation $Ro_H=Re_H Ek$.}
\label{Table:Paramo}
\end{table}

The dynamics of the fluid flows found in the celestial bodies listed in table~\ref{Table:Paramo} is further 
rendered complex by additional ingredients such as spherical geometry, multiple contributors to the density, 
compressibility, and the presence of magnetic fields. Absent such complexities, the quintessential 
paradigm for investigating rotationally influenced buoyant flows is provided by rapidly rotating 
Rayleigh-B\'enard convection (RRRBC). A large number of studies have been published on this model 
system, which is very well suited for detailed experimental, numerical and theoretical studies, 
see for example \citet{chandrasekhar1953instability,nakagawa1955theoretical,veronis1959cellular,rossby1969study,
boubnov1986experimental,zhong1991asymmetric,julien1996rapidly,knobloch1998rotating,hart2002mean,vorobieff2002turbulent,boubnov2012convection,king2012heat,stevens2013heat} and the recent review by \citet{ecke2023turbulent}. In RRRBC, gravity is often taken to be 
antiparallel to the rotation axis on the assumption that the Froude number ${Fr}_\Omega \equiv  \Omega^2 H/g$,
where $g$ is the gravitational acceleration, is small. The resulting configuration is translation-invariant and believed to be relevant to the polar regions of planets, moons and stars as well as to their interiors, although the case of misaligned gravity and rotation axis has also been studied \citep{Hathaway_Somerville_1983, julien1998strongly, julienKnoblochMilliffWerneJFM06, miquelPRF18,Novi2019,barker2020,cai2020penetrative,julien_ellison_knobloch_jfm}.
 
A major obstacle in the study of rapidly rotating convection, even in the highly
idealised setting of RRRBC, is the difficulty, to date, of attaining the extremely 
low Ekman numbers, typically $10^{-12}$ or smaller (cf. table~\ref{Table:Paramo}), that characterize 
geo- and astrophysical flows. By contrast,  current state-of-the-art DNS of the full 
Navier-Stokes-Boussinesq equations that describe RRRBC have been restricted to 
$\ekman \gtrsim 10^{-8}$ \citep{kunnen2016transition,guervilly2019turbulent,song2024direct,Song_Shishkina_Zhu_2024}, 
despite enormous computational effort. Inclusion of magnetic fields in the simulations 
to capture the capability for dynamo action, as well as spherical shell geometry, 
further limits the reported range of nearly all studies to $\ekman\gtrsim 10^{-7}$, 
${Re}_H\lesssim 10^3$ 
\citep{schaeffer2017turbulent,cooper2020subcritical,mason2022magnetoconvection,kolhey2022influence,majumder2024self,gastine2023latitudinal};
simulations down to $\ekman=10^{-10}$ are available for {\it linearised} dynamics 
only \citep{he2022internal}.
State-of-the-art laboratory experiments are likewise restricted to $\ekman\gtrsim 10^{-8}$ 
\citep{shew2005liquid,aurnou2015rotating,rajaei2017exploring,cheng2020laboratory,lu2021heat,hawkins2023laboratory,potherat2024seven} 
although the Eindhoven TROCONVEX experiment is in principle capable of achieving yet 
lower $\ekman\sim 5\times 10^{-9}$, cf.~\cite{chengGAFD18}. These accessible parameter 
values in numerical or laboratory experiments are orders of magnitude larger than the 
estimates for realistic geo- and astrophysical applications. 
On the other hand, existing reduced equations \citep{kJ98a,mS06}, known as the non-hydrostatic 
quasi-geostrophic (NHQG) equations, describe the dynamics in the limit $\ekman\to 0$, but 
their applicability is, by construction, confined to asymptotically small values of $\ekman$. 
This leaves an unexplored gap corresponding to small but finite Ekman numbers, 
precisely the regime that is most relevant for geo- and astrophysical applications. 
 
It is extremely rare to have a numerical or laboratory approach that allows one 
to span the parameter range from prototypical laboratory and DNS settings all 
the way to realistic geophysical and astrophysical settings. Remarkably, taking 
advantage of the characteristics of rapidly rotating thermal convection leads to 
a new numerical approach that relies on appropriately rescaled equations, and enables 
exploration of this very wide parameter range.

In the following section we describe the rescaled rapidly rotating 
Navier-Stokes equations (RRRiNSE) and sketch the numerical method used to solve them, 
followed in \S~\ref{sec:res} by a description of the results obtained, focusing both 
on bulk transport properties and the structure of the boundary layers at the top and 
bottom. The paper concludes in \S~\ref{sec:disc} with a discussion of the significance 
of the RRRiNSE reformulation and prospects for future work. An Appendix provides additional 
details on the flow statistics and structure.

\section{Methodology}
\label{sec:methods}
In this section, we succinctly present the RRRiNSE, an equivalent reformulation of the full
Navier--Stokes--Boussinesq equations, informed by the scaling behaviour
valid in the limit of vanishing $\ekman$~\citep{kJ98a,kJ12}. 
The RRRiNSE formulation is described in detail in a recent paper \citep{julien2024rescaled}, 
where it was first introduced and validated against existing results in the literature, 
and where the full details of the numerical implementation are given. 

In short, the RRRiNSE retain all terms in the Navier--Stokes--Boussinesq equations while 
rescaling each variable according to its leading-order asymptotic scaling behaviour valid 
for $\ekman \to 0$ in terms of the natural small parameter of rapidly rotating convection 
$\varepsilon \equiv\ekman^{1/3}$  \citep{chandrasekhar1953instability,mS06}.
In particular, the rapid rotation generates a high degree of anisotropy, which is explicitly 
incorporated in the RRRiNSE formulation: vertical length scales are non-dimensionalised by 
the layer height $H$, while horizontal length scales are significantly smaller, given 
by $\varepsilon H$. Consequently, time is measured in units of the viscous diffusion time 
based upon $\varepsilon H$. 
As shown by \citet{julien2024rescaled}, this procedure ensures an adequate preconditioning 
of the equations of motion at small but finite $\ekman$, thus facilitating DNS in this 
highly relevant but unexplored regime.

\subsection{Rescaled Navier-Stokes equation}
The convective fluid layer confined by horizontal surfaces separated by a distance $H$ 
is subjected to rotation around the vertical axis $\boldsymbol{\Omega}=\Omega\ez$ and 
uniform gravity $-g \ez$. The temperature difference between the bottom and the top 
plate is $\Delta_T>0$. In the Oberbeck--Boussinesq limit,
the fluid density varies linearly with temperature $\rho = \rho_0 [1-\alpha(T-T_0)]$ 
near a reference state $(\rho_0, T_0)$, with both the thermal expansion coefficient $\alpha$, 
the fluid kinematic viscosity $\nu$, and the temperature diffusivity $\kappa$ considered constant. 
The dimensional governing equations for the solenoidal velocity field $\boldsymbol{U}$, 
the temperature $T$ and the pressure $P$ are:
\begin{gather}
\label{eq:dimensional_boussinesq_u}
\partial_t \boldsymbol{U} + \boldsymbol{U\cdot \nabla U} 
	+ 2 \boldsymbol{\Omega} \times \boldsymbol{U}
	= -\frac{1}{\rho_0}\nabla P + \alpha g T \ez 
	+ \nu \Delta \boldsymbol{U},\qquad \boldsymbol{\nabla \cdot U} = 0\,,\\
\label{eq:dimensional_boussinesq_T}
\partial_t T + \boldsymbol{U\cdot \nabla} T = \kappa  \Delta T.
\end{gather}
As described in detail by \citet{julien2024rescaled}, the power behind the RRRiNSE formulation 
lies in their use of rotation-influenced characteristic scales for non-dimensionalisation. 
For convenience, we introduce the small parameter 
$\varepsilon \equiv \ekman ^{1/3} = \left(\nu/2 \Omega H^2\right)^{1/3}$ and measure lengths 
in units of the layer depth $H$ in the vertical, but in units of $\ell_*=\varepsilon H$ in 
the horizontal, so that $(X,Y,Z)=(\varepsilon H \widetilde{x}, \varepsilon H \widetilde{y},H\widetilde{z})$. 
In these coordinates, the rescaled anisotropic gradient 
$\boldsymbol{\nabla}=\varepsilon^{-1}H^{-1}\boldsymbol{\widetilde{\nabla}}_\varepsilon\equiv\varepsilon^{-1}H^{-1}(\partial_{\tilde{x}}, \partial_{\tilde{y}}, \varepsilon \partial_{\tilde{z}})$ 
and the Laplace operator becomes $\Delta = \varepsilon^{-2}H^{-2} \widetilde{\nabla}_\varepsilon^2$,  
where we have introduced the anisotropic diffusion operator 
$\widetilde{\nabla}_\varepsilon^2 = \partial_{\tilde{x}\tilde{x}} + \partial_{\tilde{y}\tilde{y}}+ \varepsilon^2 \partial_{\tilde{z}\tilde{z}} $. 
All tildes will be dropped in the following. Time is non-dimensionalised using a diffusive 
time-scale based on the horizontal scale rather than the layer depth alone: 
$t_*=\ell_*^2/\nu = \varepsilon^2 H^2/\nu$. Accordingly, the velocity scale is 
$U_*=\ell_*/t_*= \nu/ (\varepsilon H)$, giving the non-dimensional velocity 
$\boldsymbol{u} = \boldsymbol{U}/U_*\equiv(u,v,w)$; the pressure scale 
$P_*=\rho_0 U_*^2/\varepsilon=\rho_0 \nu^2/(\varepsilon \ell_*^2)$ is used to define 
the non-dimensional pressure $p\equiv P/P_*$. This isotropic velocity scale, combined with 
our anisotropic coordinate system, results in horizontally-dominated material derivatives: 
\begin{equation} 
    D_{t}^{\perp,\varepsilon} = \partial_t + u \partial_x + v \partial_y + \varepsilon w\partial_z
	= \partial_t + \boldsymbol{u_\perp \cdot \nabla_\perp} + \varepsilon w \partial_z.
\end{equation} 
Here, the perpendicular symbol $\perp$ in the subscript denotes the projection upon the 
horizontal plane: specifically,
$\boldsymbol{\nabla}_\perp=\left(\partial_x, \partial_y,0\right)^\top$ and
$\boldsymbol{u_\perp} = \boldsymbol{u} - w\, \ez$.
The incompressibility condition becomes:
\begin{equation}
    \boldsymbol{ \nabla_\perp \cdot u_\perp} 
    + \varepsilon \partial_z w = 0. 
    \label{eq:incomp_rescaled}
\end{equation}
The (dimensional) temperature $T$ is decomposed as the sum of three terms: an unstable 
linear stratification, a vertical profile $\Theta(z)$, and horizontal fluctuations 
$\theta(x,y,z)$ with zero horizontal mean:
\begin{equation}
        T -T_0 = \Delta_T \left[1 - z + \Theta(z) + \varepsilon \theta(x,y,z) \right],
\end{equation}
where $0\leq z \leq 1$. We thus formally embed in our decomposition one hallmark of 
rapidly rotating convection: horizontal temperature fluctuations are dominated by 
vertical variations. With these elements, the dimensionless governing equations for 
the horizontal fluctuations $(\partial_x, \partial_y) \neq (0,0)$ of momentum and 
temperature are found to be:
\begin{subequations}
\label{eq:rescaled_finitekxky}
\begin{gather}
   \label{eq:rescaled_finitekxky:momentum}
    D_t^{\perp,\varepsilon} \boldsymbol{u} + \varepsilon^{-1} \ez \times \boldsymbol{u}
	= -\varepsilon^{-1}\boldsymbol{\nabla}_\varepsilon  p 
	+ \frac{\widetilde{\rayleigh}}{\prandtl } \theta\, \ez 
	+ {\nabla}_\varepsilon^2\boldsymbol{u}\,,  \\ 
    \label{eq:rescaled_finitekxky:temperature}
    D_t^{\perp, \varepsilon} \theta  + w \left(\partial_z \Theta -1 \right) 
	= \frac{1}{\prandtl} {\nabla}_\varepsilon^2 \theta,
\end{gather}       
\end{subequations}
where $\widetilde{Ra}=\ekman^{4/3} { Ra}$, and the Rayleigh number $Ra$ is 
given by $Ra= \alpha g H^3 \Delta_T /(\nu \kappa)$ in terms of the thermal 
expansion coefficient $\alpha$ of the fluid and gravitational acceleration $g$.
By denoting the horizontal average of any quantity using an overline 
\begin{equation}
    \overline{q}^{xy} \equiv \frac{1}{L_xL_y}\int_0^{L_x}\mathrm{d}x\int_0^{L_y}\mathrm{d}y\, q(x,y,z,t)\, ,
\end{equation} 
we write separately the equations governing the vertical profiles of horizontal 
velocity $(\overline{u}^{xy}, \overline{v}^{xy})$ and temperature $\Theta$:
\begin{subequations}
\label{eq:mean}
\begin{align}
    \partial_t \overline{u}^{xy} + \varepsilon \partial_z 
    \overline{wu}^{xy} - \varepsilon^{-1}\overline{v}^{xy} & 
	= \varepsilon^{2} \partial_{zz}\overline{u}^{xy}\,, \\
        \partial_t \overline{v}^{xy} + \varepsilon \partial_z 
    \overline{wv}^{xy} + \varepsilon^{-1}\overline{u}^{xy} &
	= \varepsilon^{2} \partial_{zz}\overline{v}^{xy}\,,\\
    \varepsilon^{-2} \partial_t \Theta + \overline{ \partial_z w\theta}^{xy} & = \frac{1}{\prandtl} \partial_{zz}\Theta.
\end{align}
 \end{subequations}
Equations~(\ref{eq:incomp_rescaled}), (\ref{eq:rescaled_finitekxky}) and (\ref{eq:mean}) 
form the RRRiNSE system, which is analytically equivalent to the dimensional equations 
Eqs.~(\ref{eq:dimensional_boussinesq_u}), (\ref{eq:dimensional_boussinesq_T}) as we have 
only rescaled, but not discarded any terms in the equations. Numerically, however, these 
equations differ drastically in the sense that (\ref{eq:incomp_rescaled}), 
(\ref{eq:rescaled_finitekxky}) and (\ref{eq:mean}) can be solved accurately for very small 
values of Ekman number, whereas the poor conditioning of the unscaled equations 
(\ref{eq:dimensional_boussinesq_u}), (\ref{eq:dimensional_boussinesq_T}) renders their solution 
impossible in practice whenever $Ek\lesssim 10^{-9}$ for reasons explained 
by \cite{julien2024rescaled}.
    
\subsection{Asymptotically reduced non-hydrostatic quasi-geostrophic equations}
\label{sec:nhqg}
In the limit of rapid rotation $\ekman,\rossby \to 0$, one can derive a reduced set of 
governing equations for the flow (presented below) in a systematic and controlled way 
using asymptotic methods~\citep{kJ98a,mS06,kJ07,kJ12}. Here, we emphasize, in a quick sketch, 
how the rescaled equations~(\ref{eq:rescaled_finitekxky}) naturally lead to the reduced 
equations when the formal limit $\varepsilon \to 0$ is considered.  
Upon inspection, if the scaling employed for deriving the rescaled equations holds and 
all dimensionless variables remain $O(1)$, the leading-order terms at order $O(\varepsilon^{-1})$ 
in equation (\ref{eq:rescaled_finitekxky:momentum}) indicate geostrophic balance:
\begin{equation}
  \ez \times \boldsymbol{u}_\perp= -\boldsymbol{\nabla}_\perp  p \,,
\end{equation}
so that the pressure serves as the geostrophic streamfunction:
\begin{subequations}
\label{eq:wpressureQG}
\begin{align}
\boldsymbol{u} 
&= - \boldsymbol{\nabla}\times (p\, \ez)  + w\, \ez + O(\varepsilon)\,,\\ 
&= -\partial_y p\, \ex + \partial_x p \, \ey + w\, \ez + O(\varepsilon).
\end{align}
\end{subequations}
To obtain prognostic equations for $p$, $w$, $\theta$, $\Theta$ we substitute 
expression~(\ref{eq:wpressureQG}) into the rescaled momentum equation and project the 
result upon $\ez$, obtaining at leading order in $\varepsilon$: 
\begin{subequations}
\label{eq:reduced}
\begin{equation}
\label{eq:reduced:w}
\partial_t w + J[p,w]+ \partial_z p = 
	\frac{\widetilde{\rayleigh}}{\prandtl}\theta + \nabla_\perp^2 w,
\end{equation}
where the Jacobian $J[p,w]\equiv\partial_x p \partial_y w 
- \partial_y p \partial_x w$ indicates horizontal advection of $w$ and 
$\nabla_\perp^2\equiv\partial_{xx} + \partial_{yy}$ emphasizes that horizontal diffusion dominates. 
We complement the vertical velocity equation with a governing equation for the streamfunction 
$p$ obtained by projecting the curl of the momentum equation on the vertical. Using 
incompressibility, one proves 
$\ez\boldsymbol{\cdot\nabla}\times \left(\ez \times \boldsymbol{u}\right) = -\varepsilon \partial_z w$, 
so that the leading-order contribution of  
$\ez\boldsymbol{\cdot\nabla}\times$ (\ref{eq:rescaled_finitekxky:momentum}) yields:
\begin{equation}
 \label{eq:reduced:streamfunction}
 \partial_t \nabla_\perp^2 p + J[p,\nabla_\perp^2 p] - \partial_z w  = \nabla_\perp^4 p.
\end{equation}
Finally, one readily obtains the leading-order contribution to the governing equation for the temperature fluctuation $\theta$ and the mean temperature $\Theta$:
\begin{align}
   \label{eq:reduced:temperature}
   \partial_t \theta +J[p,\theta] + w \left(\partial_z \Theta -1 \right) &= \frac{1}{\prandtl} {\nabla}_\perp^2 \theta\,,
   \\
   \label{eq:reduced:meanTemperature}   \varepsilon^{-2} \partial_t \Theta + \overline{ \partial_z w\theta}^{xy} & = \frac{1}{\prandtl} \partial_{zz}\Theta.
   \end{align}
   \end{subequations}
Equations~(\ref{eq:reduced}) constitute the reduced Non-Hydrostatic Quasi-Geostrophic (NHQG) 
equations previously derived using a multiple-scale asymptotic expansion~\citep{kJ98a,mS06}. 
These equations filter out fast inertial waves and thin Ekman layers and govern the leading 
order geostrophically balanced flow. 

The Ekman number dependence that remains in the governing 
equation~(\ref{eq:reduced:meanTemperature}) for the mean temperature reflects the fact 
that the timescale $\tau= \varepsilon^{2}t$ associated with the evolution of $\Theta$ 
is much slower than the convective timescale. Previous studies~\citep{mS06,kJ12} 
have established that, in the limit of horizontally extended domains where a large number 
of convective cells or plumes contribute to the horizontal spatial averaging, this term can 
be omitted for the computation of global diagnostics of a statistically stationary state 
$\partial_t {\Theta}\approx 0$. This assumption can be justified based on the following 
rationale: as the horizontal extent $L_\perp=L_x=L_y$ of the domain increases, so does 
the number of independent plumes $N_p$ that feed back into the temperature profile. 
The central limit theorem then suggests that, in the ergodic limit (time and ensemble 
averages behave similarly), the amplitude of temporal fluctuations of the vertical 
temperature profile $\Theta(z)$ should decrease like $1/\sqrt{N_p}\approx 1/L_\perp$, 
justifying the neglect of the term $\partial_t\Theta$. In the context of the RRRiNSE 
formulation, this assumption has been successfully validated and has proved to shorten 
the transient that leads to the stationary state~\citep{julien2024rescaled}. 
Thus in all simulations of the RRRiNSE (\ref{eq:incomp_rescaled}), (\ref{eq:rescaled_finitekxky}) 
and (\ref{eq:mean}) and the NHQG equations (\ref{eq:reduced}) described in this paper, 
$\partial_t\Theta$ is set to zero, as is expected to be the case for a sufficiently large 
horizontal extent $L_\perp$ of the domain. 

\subsection{Details on the numerical method}
\label{sec:numerics}
We compute solutions to both the rescaled equations (\ref{eq:incomp_rescaled}), 
(\ref{eq:rescaled_finitekxky}) and (\ref{eq:mean}) and the reduced equations (\ref{eq:reduced}) 
using the pseudo-spectral code Coral~\citep{miquel2021coral}, a flexible platform for 
solving partial differential equations in a Cartesian domain. We assume periodicity 
in $(x,y)$ and thus decompose the problem variables using a Fourier basis in the horizontal, 
whereas a Chebyshev expansion is used in the vertical in order to impose boundary conditions on 
the bounding surfaces $z=0,1$ using a Galerkin basis recombination. Differential operators 
are discretised using the quasi-inverse method 
\citep{clenshaw1957numerical,greengard1991spectral,kJ09} to optimize memory usage and accuracy. 
The resolution was varied from $192$ grid points in each direction (corresponding to $128$ 
alias-free modes) at the lowest values of $\widetilde{Ra}$, up to $768$ grid points 
($512$ Chebyshev modes) in the vertical direction and $576$ grid points ($384$ Fourier modes) 
in the horizontal directions at larger $\widetilde{Ra}$. The time-marching relies on 
the family of stable semi-implicit Runge--Kutta schemes introduced by~\citet{Ascher97}, 
specifically their four-stage, third-order scheme, known as ARS443.

For some of the runs presented here, we have used low-amplitude noise as the initial condition, 
and waited for the convective turbulent flow to settle into a statistically stationary state 
after a transient period. However, for strongly supercritical flow, this transient may prove 
much more turbulent than the stationary state itself. Hence, in some cases, we resort to an 
alternative: instead of low-amplitude noise, the initial condition for a given run was taken 
from the solution computed at a neighboring position in parameter space. This procedure 
facilitates the computation of the transient. 

Once the transient has subsided, we compute time averages of different  diagnostic quantities 
reported here. For instance, the barotropic vertical vorticity displayed in 
figure~\ref{fig:vorticity_bulk} is defined as:
\begin{equation}
\overline{\omega_z}(x,y) = \int_0^1\omega_z\,\mathrm{d}z\,,
\end{equation}
where the vertical vorticity is given by $\omega_z=\partial_x v-\partial_y u$. 
The Nusselt number is computed in terms of a volume average as 
$Nu(t)  -1 = \prandtl\langle w \theta\rangle_{\rm vol} $, where the 
volume average is given by 
\begin{equation}
\langle \cdot \rangle_{\rm vol} 
	= \overline{\overline{(\cdot)}^{xy}} 
	= \frac{1}{L_xL_y}\int_0^{L_x}\mathrm{d}x\int_0^{L_y}\mathrm{d}y\int_0^1 \left( \cdot \right) \,\mathrm{d}z\,,
\end{equation} 
and the small-scale Reynolds number based on the vertical velocity and the horizontal 
length scale is given, in the non-dimensional RRRiNSE formulation, by 
$Re(t) \equiv \sqrt{\langle w^2 \rangle_{\rm vol}}$. In addition, we denote the 
combined space and time average by:
\begin{equation}
\langle \cdot \rangle 
	= \frac{1}{t_2-t_1} \int_{t_1}^{t_2}\mathrm{d}t \frac{1}{L_xL_y}
	\int_0^{L_x}\mathrm{d}x\int_0^{L_y}\mathrm{d}y\int_0^1 \left( \cdot \right) \,\mathrm{d}z . 
\end{equation}

\section{Results}
\label{sec:res}
Here, we present the first DNS of RRRBC at realistically small Ekman numbers for planetary, 
satellite and stellar interiors, using the RRRiNSE formulation, and analyze the physics 
of the flow in this extreme parameter range. Our extensive set of simulations was performed 
using the pseudospectral solver Coral \citep{miquel2021coral} for a periodic domain of 
horizontal extent $10\ell_c$, where $\ell_c\approx 4.82\ekman^{1/3}$ is the critical wavelength 
in the limit $\ekman \to 0$~\citep{chandrasekhar1953instability}. Specifically, we consider a 
fluid with $\sigma =1$ subject to stress-free boundaries (for simplicity) at top and bottom where 
constant temperatures are imposed, and periodic boundary conditions in the horizontal directions.
We explore unprecedentedly small but finite Ekman numbers, as low as $ Ek=10^{-15}$, values 
that are more than six orders of magnitude smaller than those previously attainable in DNS or 
in the laboratory.
\begin{figure}
\centering
\begin{overpic}[width=\textwidth]{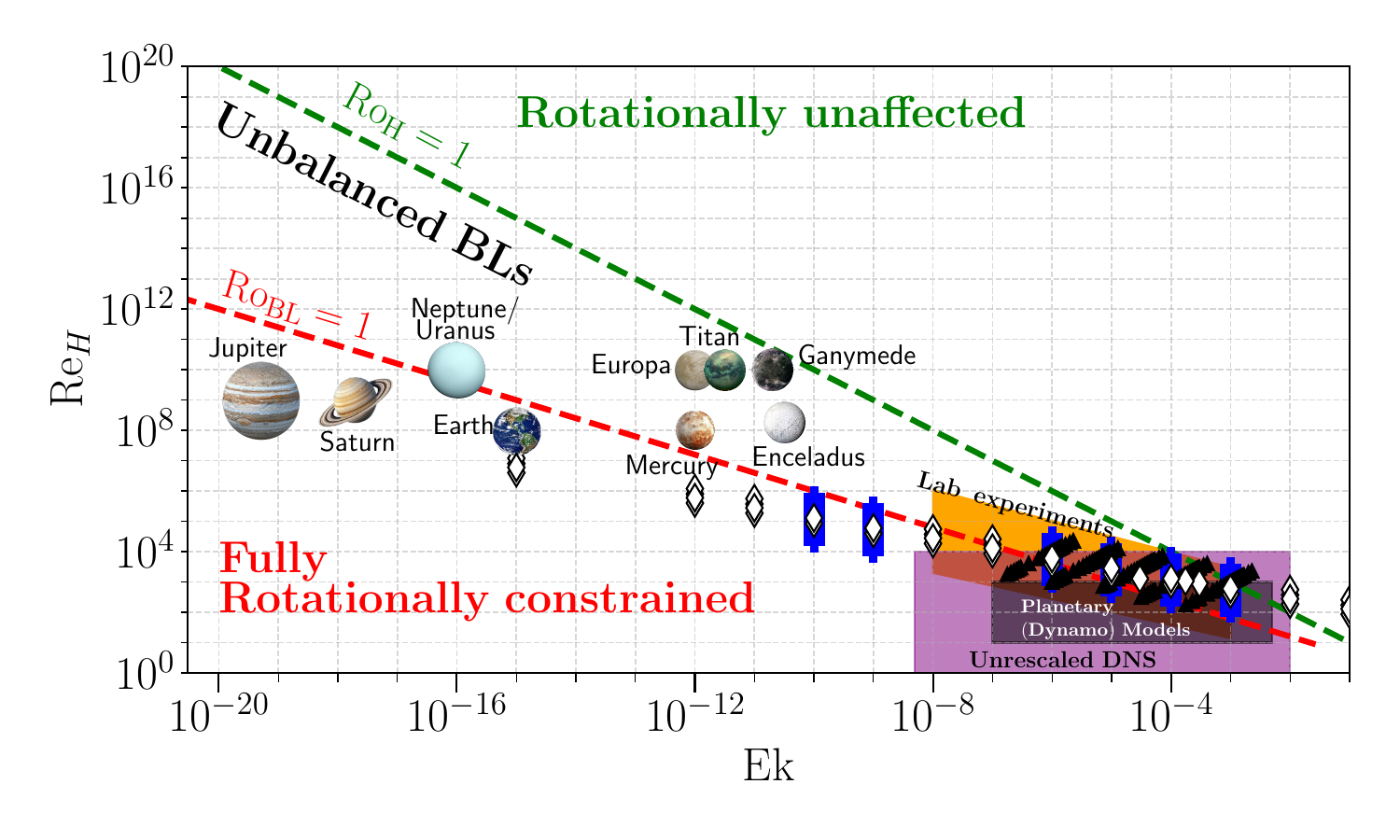}
	\put (0,53.5) {\Large $(a)$}
\end{overpic}
 \vspace{-0.72cm}\\ 
\begin{overpic}[width=1\textwidth]{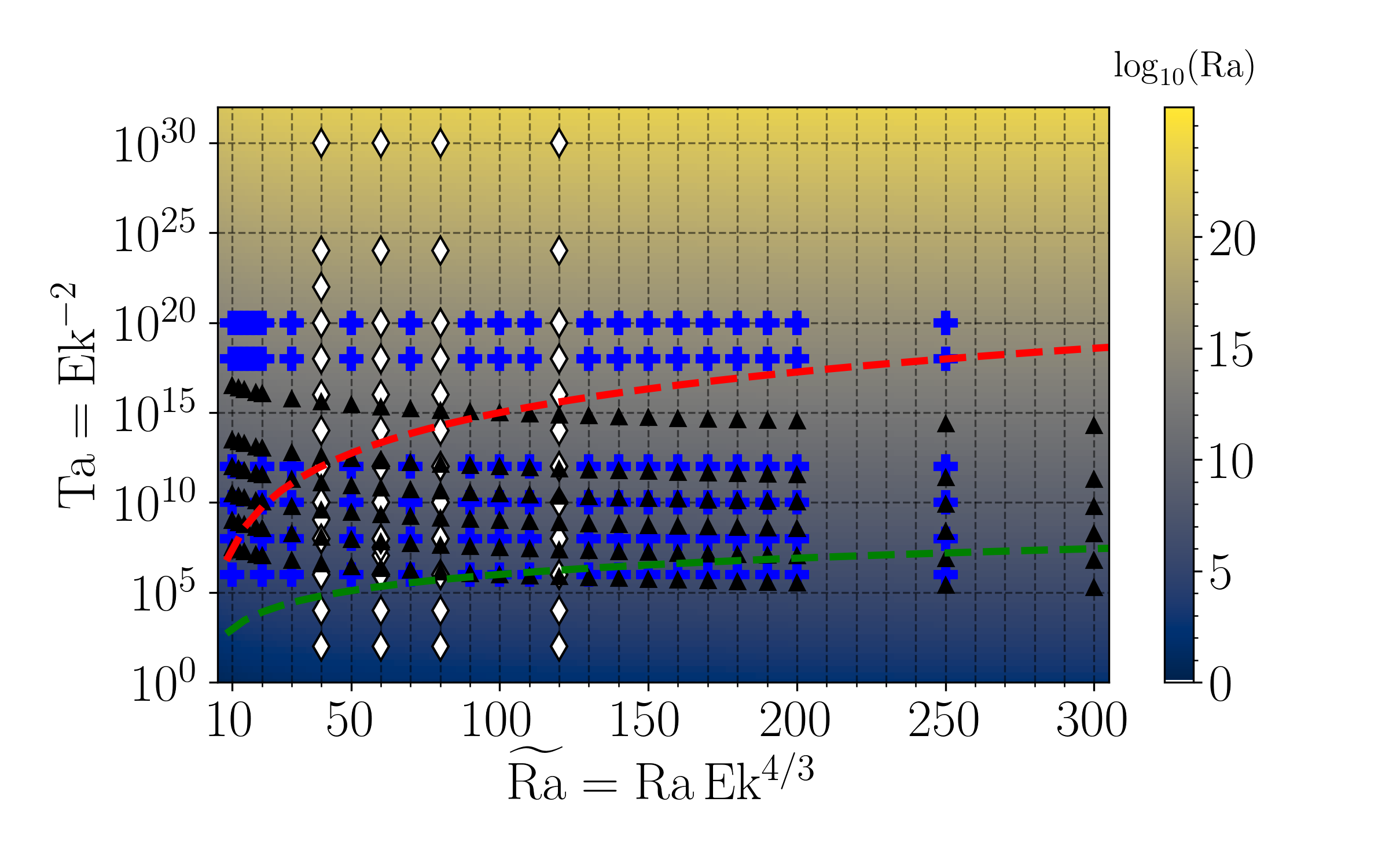}
	\put (0,52) {\Large $(b)$}
\end{overpic}
\vspace{-1.1cm}
 \caption{\raggedright 
Overview of the parameter space. Panel $a$: estimates of the non-dimensional parameters 
for different celestial objects (see table~\ref{Table:Paramo}) shown in the plane spanned 
by the Ekman and Reynolds numbers. The parameters reached in previous laboratory experiments
and simulations are indicated by shaded regions. The dashed lines indicate regime boundaries
discussed in the text: the green dashed line corresponds to the bulk Rossby number 
$Ro_H=1$ while the red dashed line corresponds to the local Rossby number 
$Ro_{\rm BL}=1$ in the thermal boundary layers (BLs). These lines are given by 
$\reynolds_H = 1/\ekman$ and $Re_H=Ek^{-3/5}$, respectively.
The regime bounded by these two dashed lines is characterised by unbalanced boundary
layers and a balanced bulk flow.
The symbols refer to the simulations summarised in panel $b$: $\widetilde{Ra}=const.$ 
(white diamonds), $Ra=const.$ (black triangles) and $\ekman=const.$ (blue plus signs). 
Panel $b$: overview of simulations (with symbols identical to panel $a$) in the plane spanned 
by the reduced Rayleigh number $\widetilde{Ra}=Ra \, Ek^{4/3}$ and the Taylor number 
$Ta=Ek^{-2}$. Dashed lines represent regime boundaries corresponding to those in panel $a$.}
 \label{fig:overview_parameter_space}
\end{figure}

\subsection{Overview of parameter space}
Figure~\ref{fig:overview_parameter_space}$a$ shows the parameter space explored in our simulations. 
The top panel shows the estimates of these parameters for planetary interiors and oceans on icy 
satellites given in table~\ref{Table:Paramo} in the plane spanned by the Ekman number $\ekman$ and 
the bulk Reynolds number $\reynolds_H$, together with blue circles showing the values reached in
our simulations, indicating that we attain realistically small $\ekman$ equal to the estimated 
values for the outer core of the Earth ($\ekman\approx 10^{-15}$) and the convective zone of 
the Sun (not shown, cf. table~\ref{Table:Paramo}). We further indicate the approximate range 
of parameters reached in previous numerical and laboratory studies, which are restricted to 
higher $\ekman$. Dashed lines highlight important regime boundaries where the Rossby number 
$\rossby$ equals unity in different parts of the flow. The green dashed line is an estimate of 
where the bulk of the flow is expected to become rotationally supported as the rotation rate 
increases, namely $\rossby_H=1$, i.e., $\reynolds_H = 1/\ekman$. In contrast, the rotation 
rate required to generate rotational support in the thermal boundary layers at the top and 
bottom is much higher. This rotation rate is determined by the condition that the local 
Rossby number in the thermal boundary layer $Ro_{BL}=1$ (red dashed line). As shown by 
\cite{julien2012heat} this condition is satisfied when the convective Rossby number associated 
with the bulk flow $Ro_{conv}=Ek^{1/5}$. With velocity following the rotational free-fall scale, 
$U=\alpha g \Delta_T/ (2\Omega)$, we see that $Re_H=Ro_{conv}^2/Ek$, 
cf.~\cite{julien2024rescaled}, and hence that $Re_H=Ek^{-3/5}$ (red dashed line in 
figure~\ref{fig:overview_parameter_space}$a$).

In rapidly rotating flows, higher values of $Ra$ are required to trigger convection as a 
consequence of the Taylor-Proudman constraint \citep{proudman1916motion}. Specifically, 
the critical Rayleigh number for convection in the limit of small $\ekman$ is proportional 
to $\ekman^{-4/3}$ as shown by \citet{chandrasekhar1953instability}. Therefore, it is natural 
to introduce $\widetilde{Ra}\equiv RaEk^{4/3}$, the \textit{reduced} Rayleigh number, which 
measures the supercriticality of rapidly rotating convection. Panel $b$ of 
figure~\ref{fig:overview_parameter_space} shows an alternative representation of the parameter 
space in terms of the reduced Rayleigh number $\widetilde{Ra}$ and the Taylor number 
$\taylor=\ekman^{-2}$. We reach Taylor numbers up to $10^{30}$ and reduced Rayleigh numbers 
up to $300$, corresponding to rapidly rotating and highly turbulent flows. 
The associated bare Rayleigh number $\rayleigh$, indicated by colors in the background, 
ranges over a similarly wide interval from $O(10^3)$ to very large values in excess of $10^{20}$.

This wide array of numerical simulations allows us to probe, for the first time in DNS, 
the physics of rotating convection from moderate Ekman numbers down to the fully rotationally 
constrained regime,  previously only accessible in asymptotically reduced 
equations~\citep{mS06,julien2012heat} within the $Ek\to 0$ limit. 

\begin{figure}
    \centering
    \begin{overpic}
    [width=\textwidth]{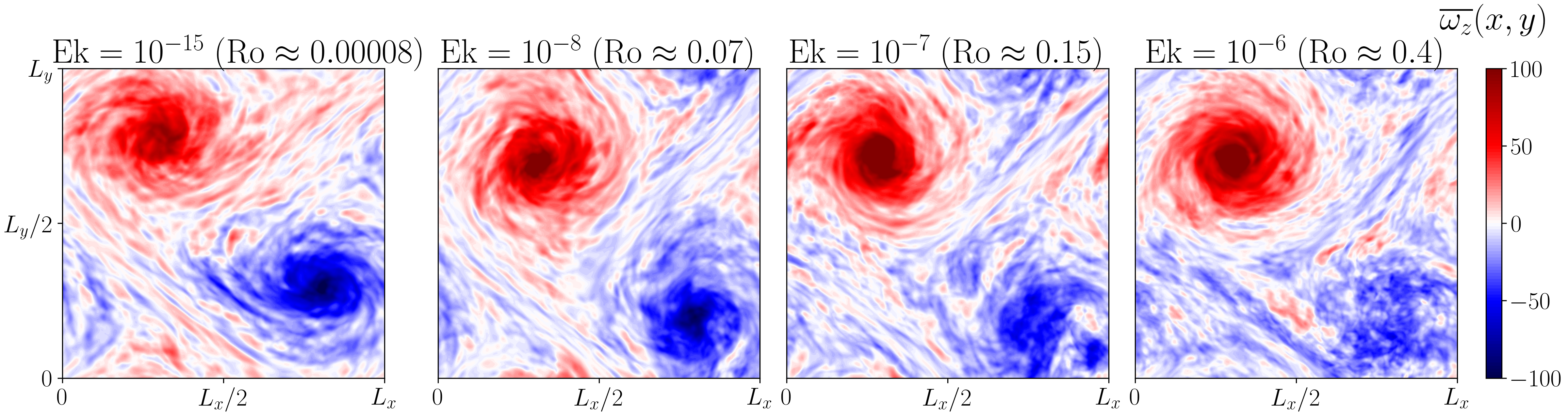}
    \put (3,26) { $(a)$ \hspace{2.7cm}$(b)$\hspace{2.6cm} $(c)$ \hspace{2.6cm} $(d)$}  
    \end{overpic}
	$(e)$\hspace{3cm}\text{ }\\
\includegraphics[width=0.35\linewidth]{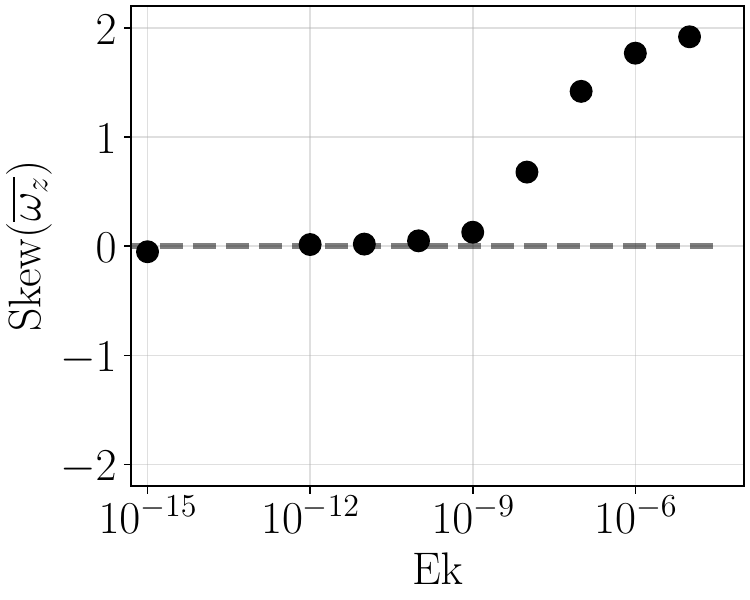}
    \caption{\raggedright
Top row: vertically averaged (i.e., barotropic) vertical vorticity field 
$\overline{\omega_z}(x,y)$ at $Ek=10^{-15}$ (panel $a$), $10^{-8}$ (panel $b$), $10^{-7}$ 
(panel $c$), $10^{-6}$ (panel $d$) and $\widetilde{\rayleigh} = 120$ (all panels). Panel $e$: 
skewness of the barotropic vorticity field versus $Ek$ indicating a strong cyclone--anticyclone 
asymmetry for $Ek\gtrsim 10^{-9}$, and an approximate (statistical) cyclone--anticyclone symmetry 
for $Ek \lesssim 10^{-9}$.}
    \label{fig:vorticity_bulk}
\end{figure}
\subsection{Flow morphology -- bulk and boundary layers}
An important feature of rapidly rotating flows is the formation of large-scale flow structures 
(vortices or jets), associated with spectral condensation of kinetic energy at the domain scale 
as a result of an inverse energy cascade \citep{favier2014inverse,guervilly2014large,aR14}. 
This behaviour, which is robust for rotating convection in different domain geometries 
\citep{guervilly2017jets,julien2018impact,lin2021large}, is also observed more broadly in 
anisotropic turbulent flows, such as stably stratified rotating turbulence, turbulence subject 
to strong magnetic fields or geometric confinement to a thin-layer geometry, among 
others \citep{alexakis2018cascades,alexakis2023quasi,van2024phase}.
In the case of a periodic, horizontally square Cartesian domain, the condensate in the limit 
of asymptotically large rotation $\ekman \to 0$ takes the form of a large-scale vortex dipole, 
first observed in the reduced equations \citep{kJ12,aR14}, resembling findings in two-dimensional 
turbulence \citep{boffetta2012two}. In contrast, direct numerical simulations of 
rotating convection at previously attainable moderately small Ekman and Rossby numbers reveal 
an asymmetry between cyclones and anticyclones, namely a preference for strong cyclonic
structures and diffuse, weaker anticyclones \citep{favier2014inverse}. While a dominance of 
cyclones over anticyclones is also observed in stably stratified flows at moderately small 
Rossby numbers~\citep{roullet2010cyclone,gallet2014scale}, the transition from the moderately 
small Rossby number regime to very small $\ekman$ (relevant to the celestial bodies listed in 
table~\ref{Table:Paramo}), where the cyclone--anticyclone symmetry is expected to be restored, 
could not previously be reached in RRRBC. 

To address this question, we illustrate and analyze in figure~\ref{fig:vorticity_bulk} how the 
vertically averaged (i.e., barotropic) vertical vorticity $\overline{\omega_z}(x,y)$ (see 
Sec.~\ref{sec:methods}) changes as $\ekman$ varies for a fixed supercriticality 
$\widetilde{\rayleigh}=120$. At moderate rotation rates ($\ekman=10^{-6}$ and $\rossby \approx 0.4$,
panel $(d)$), a striking asymmetry exists between the strong, coherent cyclone and a weak, 
incoherent anticyclonic region. Here, the \textit{a posteriori} Rossby number $Ro$ is computed 
as the root mean square of the ratio 
$\left\|\boldsymbol{u \cdot \nabla u}_\perp\right\|/(2\Omega \left\|\boldsymbol{u}_\perp\right\|)$, 
where $\boldsymbol{u}_\perp = (u,v,0)$. As rotation increases (top row of 
figure~\ref{fig:vorticity_bulk}, right to left), this asymmetry decreases but remains visible 
for $\ekman = 10^{-8}$ ($\rossby \approx 0.07$, panel $b$), a value close to the current state 
of the art for RRRBC studies. The anticipated cyclone--anticyclone symmetry present in the 
asymptotic rapidly rotating limit $\ekman \to 0$ based on the NHQG equations is in fact 
observed for $\ekman = 10^{-15}$ ($\rossby \approx 8\times10^{-5}$, panel $a$). To quantify 
this qualitative observation, we display in figure~\ref{fig:vorticity_bulk}$(e)$ the skewness 
of the barotropic vertical vorticity: 
\begin{equation} 
\mathrm{Skew}(\overline{\omega_z}) = 
 \left. 
 \left\langle \left( 
   \overline{\omega_z} - \left\langle \overline{\omega_z} \right \rangle 
             \right)^3 \right \rangle 
\middle/
 \left\langle \left( 
   \overline{\omega_z} - \left\langle \overline{\omega_z} \right \rangle 
             \right)^2 \right \rangle^{3/2}. 
\right.
\end{equation}
A finite positive value, indicating the prevalence of strong cyclones, is obtained 
for $\ekman \gtrsim 10^{-9}$, whereas an approximately vanishing skewness is observed at 
smaller Ekman numbers, indicating a restored (statistical) cyclone--anticyclone symmetry, 
with a relatively sharp transition between the two regimes. 
\begin{figure}
\centering
\begin{overpic}
	[width=\textwidth, trim={1cm, 9cm, 2.1cm, 9cm}, clip] {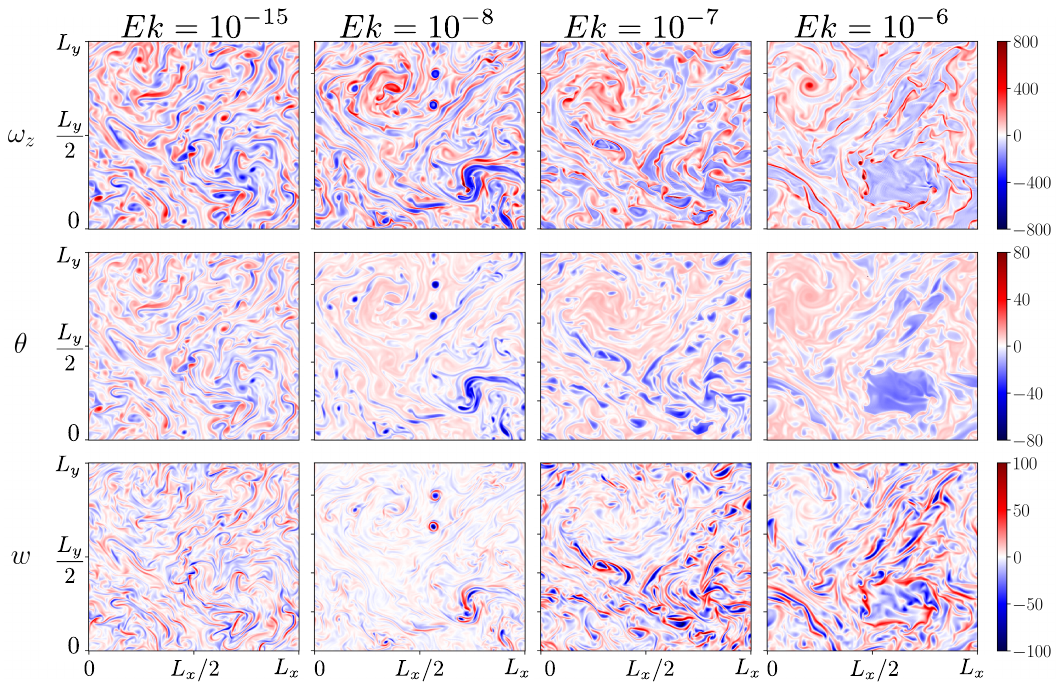} \put (-2,62) {\large $(a)$}
\end{overpic}  \\
	\vspace{2mm}
    \begin{overpic}
        [width=\textwidth]{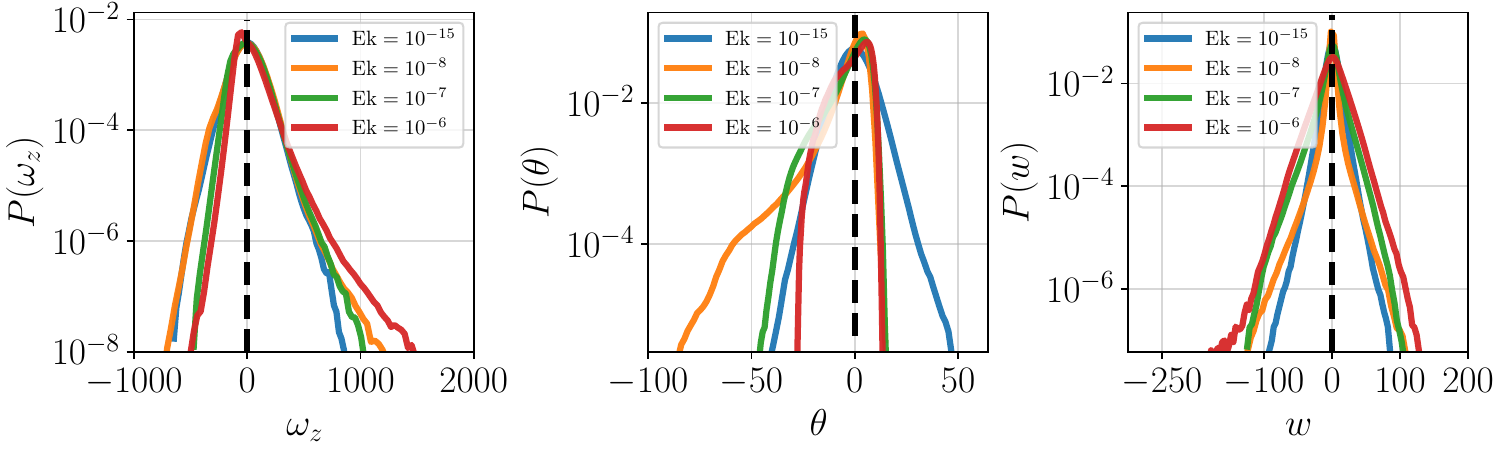}
	    \put (-2,28) {\large$(b)$}
    \end{overpic}
    \begin{overpic}
	    [width=\textwidth]{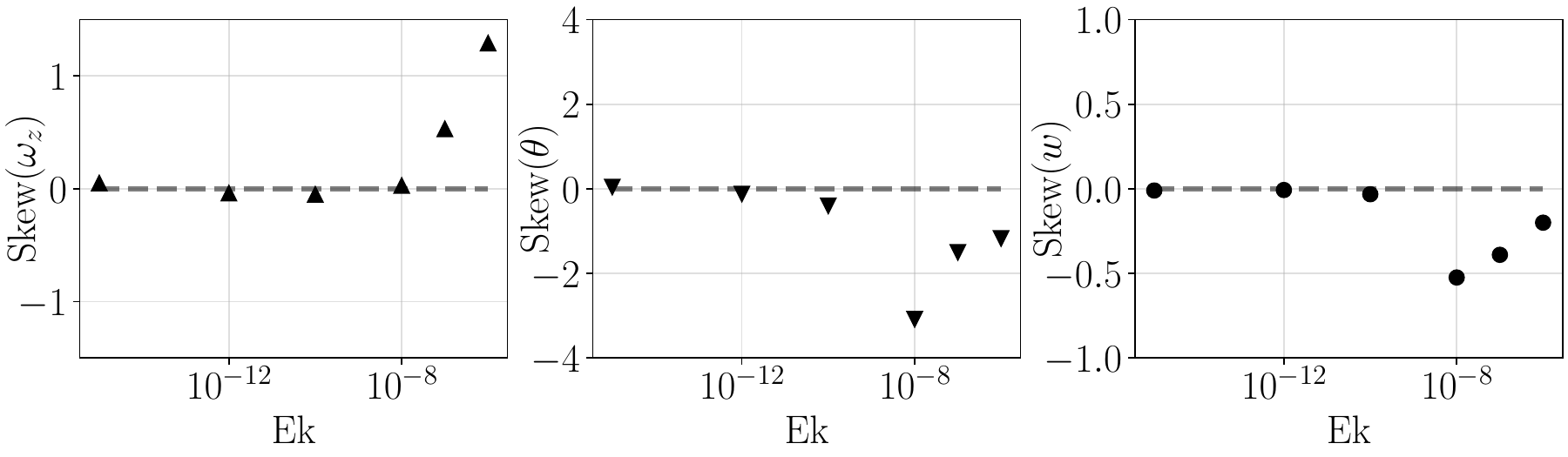} \put (-2,27) {\large $(c)$}
     \end{overpic}

\caption{\raggedright
Panel $a$: overview of boundary layer flow morphology in terms of the vertical 
vorticity $\omega_z$, temperature perturbation $\theta$ and vertical velocity $w$ at 
different $\ekman$ and $\widetilde{\rayleigh} = 120$. 
Panel $b$: histograms of 
$\omega_z,\theta,w$ at different $\ekman$. Near-Gaussian statistics are observed 
at $\ekman = 10^{-15}$ with larger skewness at larger $\ekman$. 
Panel $c$: skewness of $\omega_z,\theta,w$ versus $\ekman$. All results shown are 
obtained at $z=\delta_{\omega_z}$ (top of the momentum boundary layer near the bottom boundary) 
	defined in figure~\ref{fig:width_of_thermal_and_momentum_bls}$(a)$. \label{fig:flow_visualisation_combined} }
\end{figure}

In addition to the barotropic component of the flow, which informs us about the bulk flow 
structure, the flow inside the thermal boundary layer near the stress-free, 
constant-temperature boundaries at the top and bottom is of great importance. 
Figure~\ref{fig:flow_visualisation_combined}$(a)$ shows visualizations of the flow inside 
the bottom thermal boundary layer (at height $z=\delta_{\omega_z}$, defined thereafter, 
cf. figure~\ref{fig:width_of_thermal_and_momentum_bls}) at $\widetilde{Ra}=120$ for 
four distinct Ekman numbers $\ekman =10^{-15},10^{-8},10^{-7},10^{-6}$ in terms of three fields: 
the vertical vorticity $\omega_z$, the temperature perturbation $\theta$ away from the 
horizontally averaged temperature profile and the vertical velocity $w$ at the top of the 
(bottom) boundary layer, where the flow transitions between qualitatively distinct regimes. 
Panel $b$ shows histograms of each field at the same Ekman numbers, while panel $c$ displays 
the associated statistical skewness versus $\ekman$. At very low Ekman numbers, 
$Ek=10^{-15}$, where the bulk flow features two large-scale vortices and displays 
an approximate statistical cyclone--anticyclone symmetry, the same symmetry is obeyed 
in the boundary layer as evidenced by the approximately vanishing skewness not only of $\omega_z$, 
but also $\theta$ and $w$ (see also Supplementary Movies 1-3). A near perfect correlation 
can be seen between the vertical vorticity and $\theta$. We mention that the RRRiNSE formulation 
remains numerically stable even when the Ekman number is reduced further, as far as 
$Ek=10^{-24}$; the corresponding flow fields obtained from DNS are shown in Appendix~\ref{sec:app_C}. 

As the Ekman number increases, the approximate symmetry of the histograms persists up 
to $Ek\approx 10^{-9}$ (not shown). At $Ek\approx 10^{-8}$, where cyclone--anticyclone 
asymmetry and cyclone dominance emerges in the barotropic mode, the boundary layer flow 
morphology is also drastically altered: in addition to the trace of the strong large-scale 
cyclone, similarly strong, albeit short-lived, anticyclonic vortices emerge in the boundary 
layer, whose anticyclonic core is seen in some cases to be surrounded by a cyclonic shield 
(see also Supplementary Movie 4). These strong vortices with anticyclonic cores have a clear 
signature in the temperature fluctuation $\theta$ as strong, cold perturbations (in the bottom 
boundary layer shown here, see Supplementary Movie 5), and also in the vertical velocity $w$ 
which shows descending flow inside the vortex core, surrounded by rising fluid motion 
(see Supplementary Movie 6). As a result, the statistics of $\theta$ and $w$ are skewed towards 
negative values at this $Ek$. It is important to stress that the structures displayed in 
figure~\ref{fig:flow_visualisation_combined} and discussed above are confined within the 
boundary layer, and do not extend into the bulk flow. To the best of our knowledge, the strong, 
partially shielded, short-lived anticyclonic vortices in the boundary layer with stress-free 
boundaries are described here for the first time, although shielded structures have been 
previously observed in various studies of rotating convection 
\citep{mS06,kJ12,sS14,song2024direct}. We also note that similar structures were predicted 
based on Ekman layer theory for no-slip boundary conditions by \citet{julien1996rapidly} in 
response to a Gaussian temperature perturbation. As $\ekman$ is increased to $10^{-7}$ and $10^{-6}$, 
cyclones clearly dominate the flow in the boundary layer (as in the bulk), while the strong 
localised anticyclones become more diffuse, leading to a significant skewness in the statistics 
of $\omega_z$. The responses in $\theta$ and $w$ remain visible in the boundary layer, 
particularly outside the large-scale cyclone, although their skewness decreases with further 
increase in $Ek$. 
\begin{figure}
\centering
\begin{overpic}
	[width=0.495\textwidth]{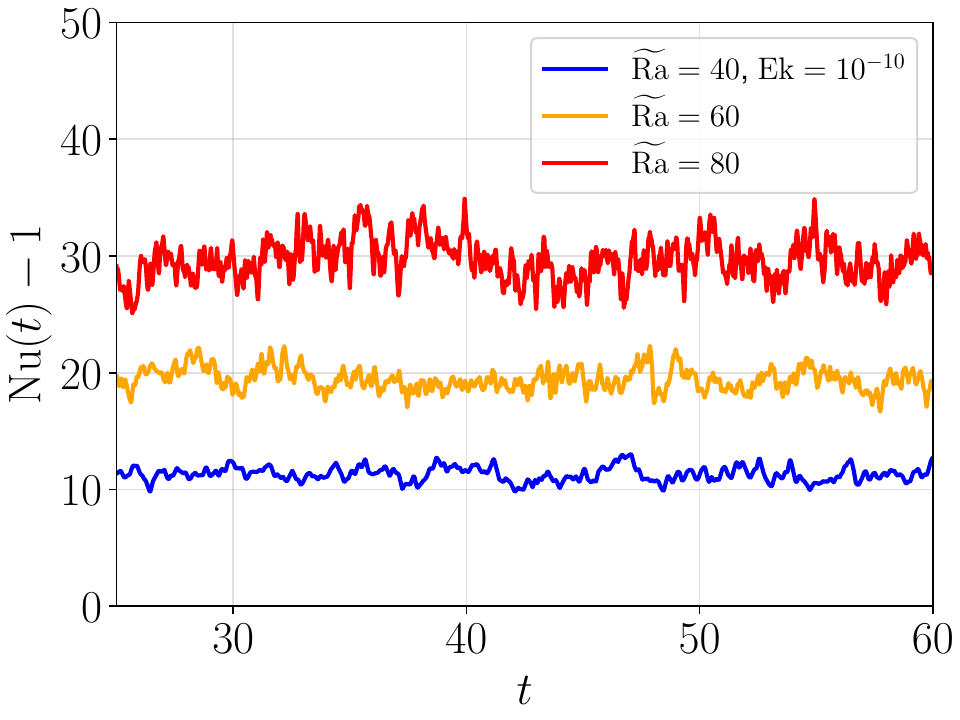} \put (0,70) {$(a)$}
\end{overpic}
\begin{overpic}
	[width=0.4845\textwidth]{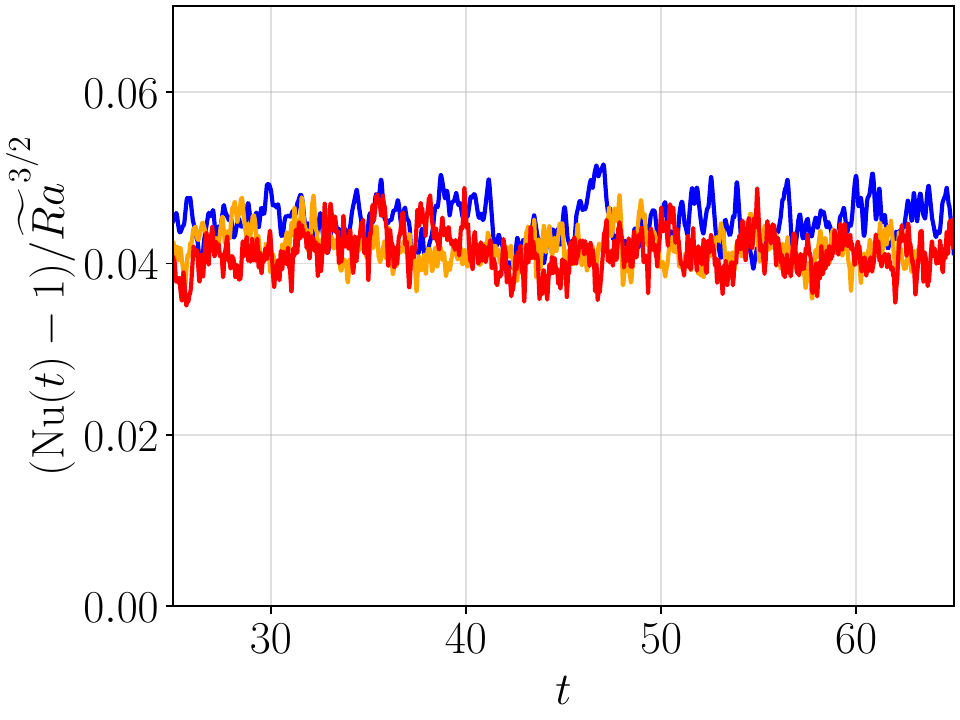}\put (0,70) {$(b)$}
\end{overpic}
\begin{overpic}
[width=0.485\textwidth]{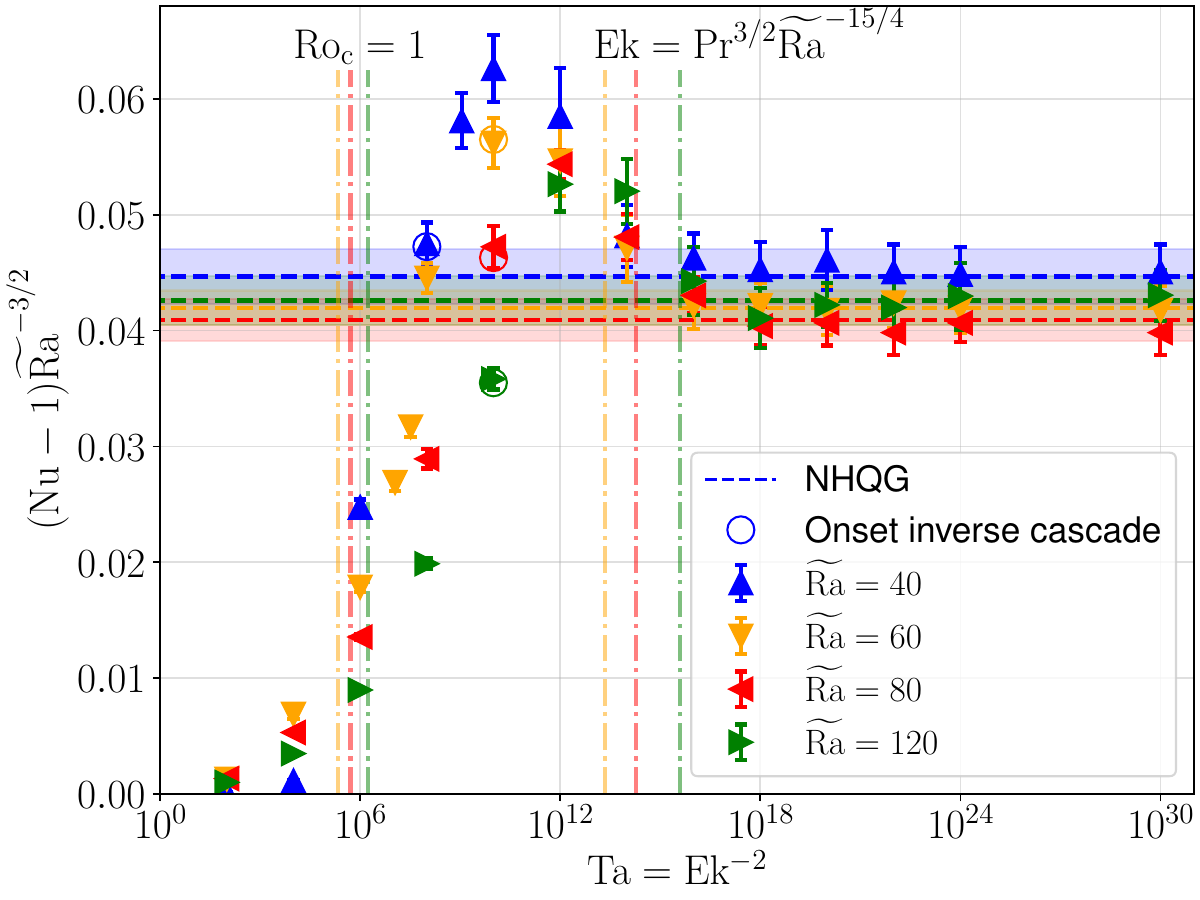} 
\put (0,70) {$(c)$}
\end{overpic}
\begin{overpic}
	[width=0.495\textwidth]{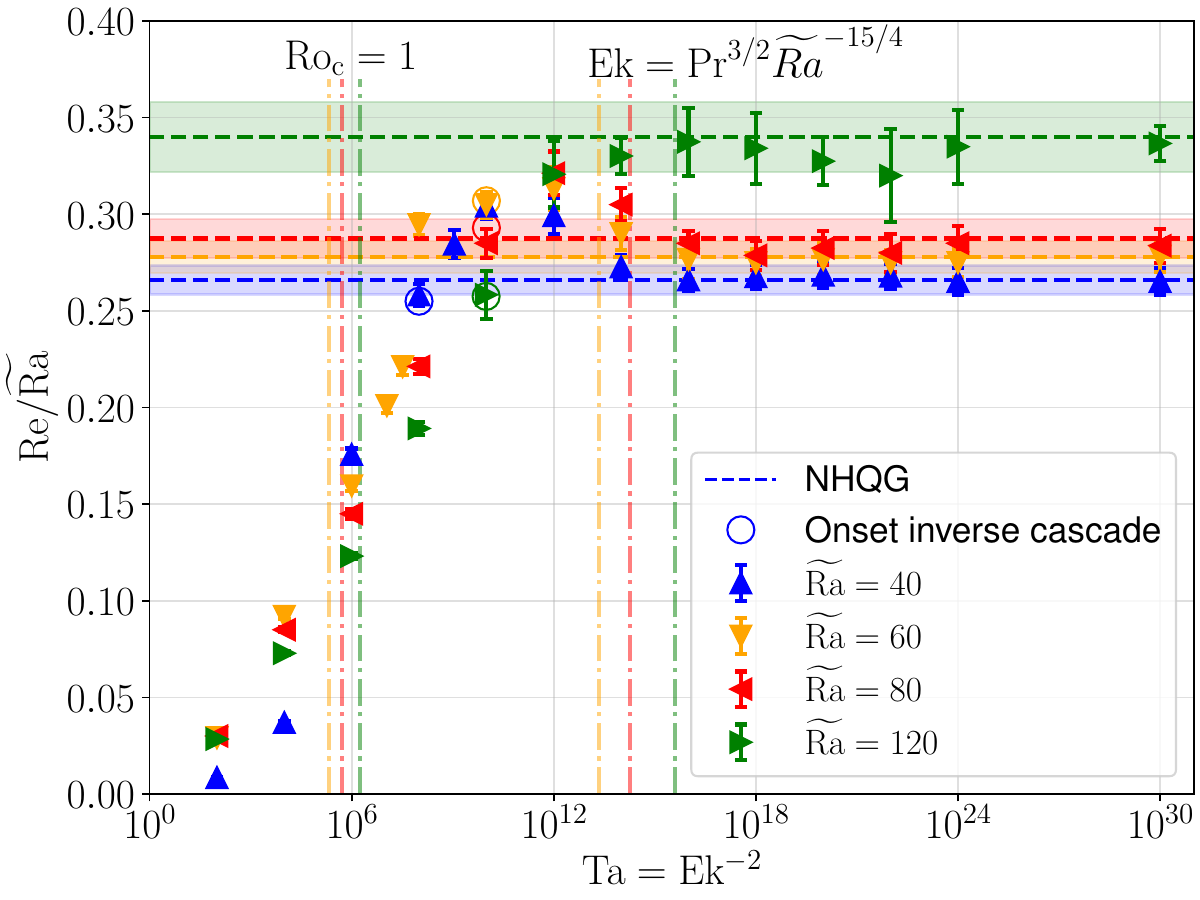} \put (0,70) {$(d)$}
\end{overpic}
\begin{overpic}
	[width=0.485\textwidth]{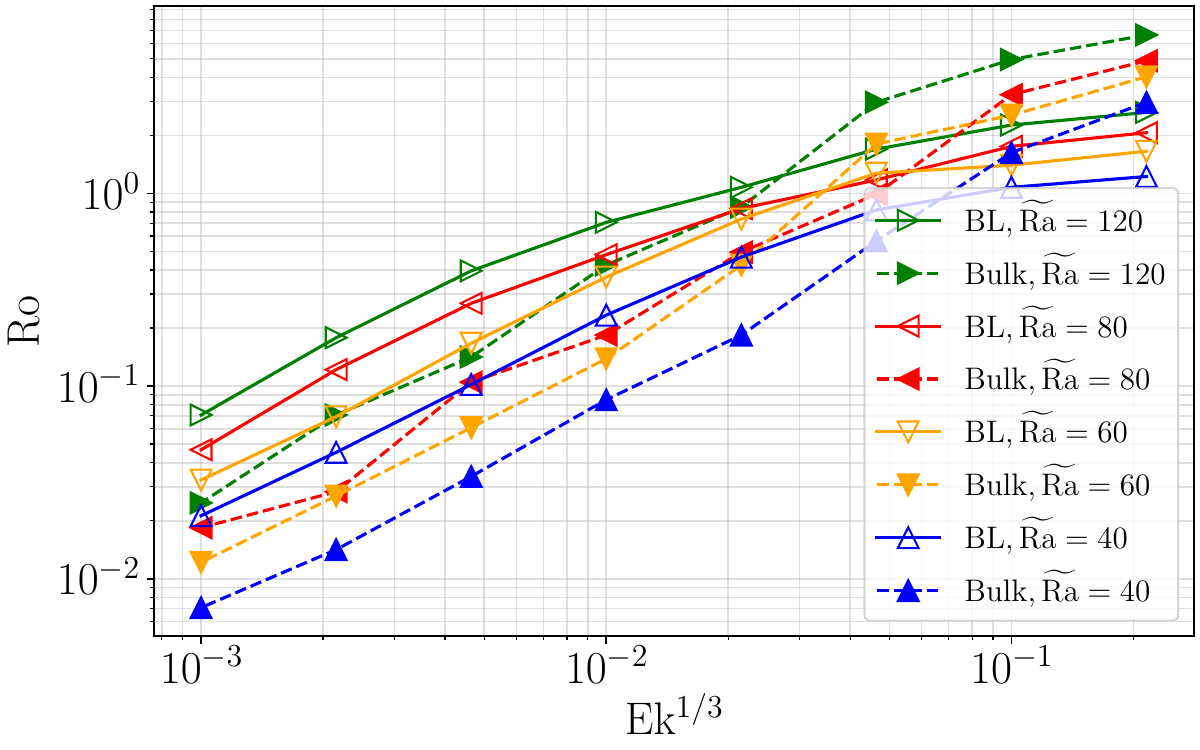} \put (0,60) {$(e)$}
\end{overpic}
\begin{overpic}
	[width=0.485\textwidth]{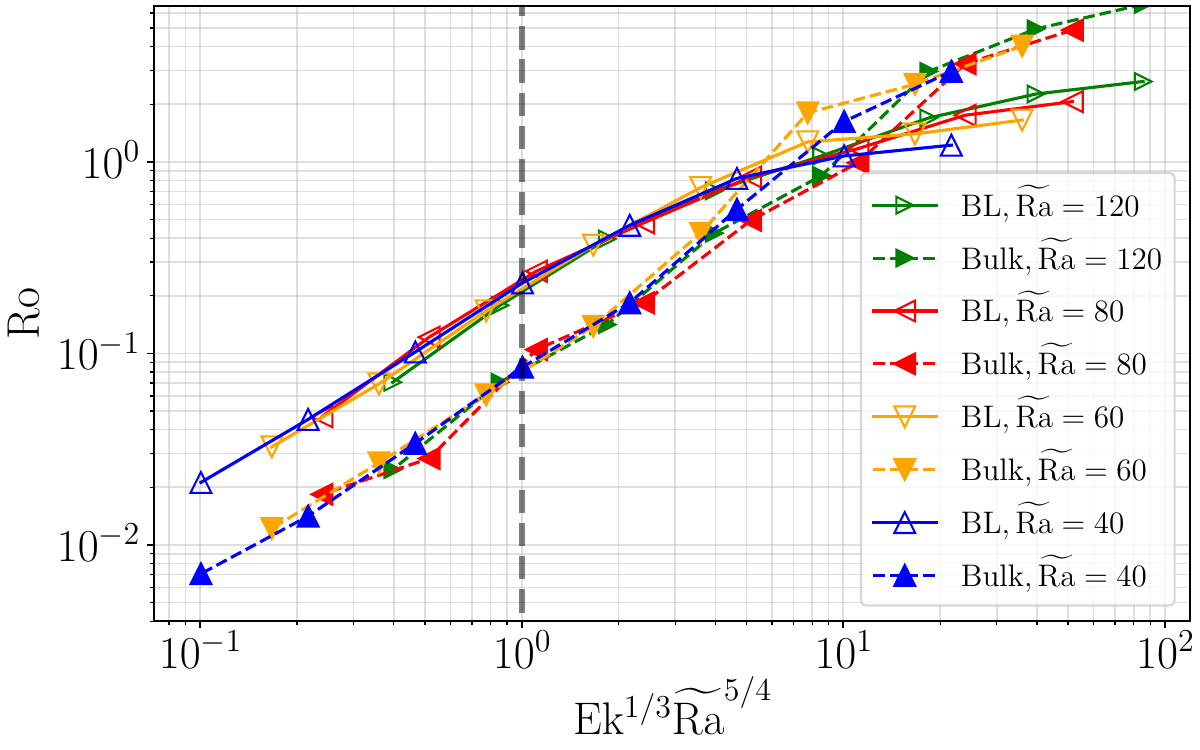} \put (0,60) {$(f)$}
\end{overpic}
\caption{\raggedright
Panel $a$: sample time series of $Nu-1$ at $Ek = 10^{-10}$ for $\widetilde{Ra}=40,60,80$ in 
the statistically stationary state, revealing turbulent fluctuations about a well-defined 
mean value. Both mean and amplitude of fluctuations increase with $\widetilde{Ra}$. 
Panel $b$: same data, collapsed by rescaling the $y$ axis by $\widetilde{Ra}^{3/2}$. 
Panel $c$: temporally averaged $Nu-1$ in the stationary state, compensated by 
$\widetilde{Ra}^{3/2}$, with error bars indicating standard deviation, vs $Ta=\ekman^{-2}$. 
$Nu$ increases with $Ta$ at fixed $\widetilde{Ra}$ up to a maximum, beyond which it 
decreases to converge to a $Ta$-independent constant, which agrees with the value obtained in 
the NHQG equations at the same $\widetilde{Ra}$ (horizontal dashed line), within one 
standard deviation (shaded area). Vertical dashed lines (same color scheme) correspond to 
$Ro_{conv}=1$ and $Ro_{BL}=1$. Panel $d$: corresponding data for the small-scale Reynolds 
number $Re\equiv \sqrt{\langle w^2\rangle }$, compensated by $\widetilde{Ra}$, show a 
similar structure with an overshoot and eventual convergence to the NHQG prediction at 
large $Ta$. Panel $e$: \textit{a posteriori} Rossby numbers versus $\ekman^{1/3}$ for 
the bulk and the boundary layer, both of which exhibit collapse in panel $f$ when shown 
versus $\ekman^{1/3} \widetilde{Ra}^{5/4}$.\label{fig:Nu_Re_Ro} 
}
\end{figure}

\subsection{Flow statistics}
Going beyond the visualizations shown in figure~\ref{fig:flow_visualisation_combined}, we 
display in figure~\ref{fig:Nu_Re_Ro} key quantitative information on the flow in terms of 
important non-dimensional numbers. We define the non-dimensional heat flux across the layer, 
measured by the Nusselt number $Nu-1\equiv \sigma \langle w \theta \rangle$ where 
$\langle \cdot \rangle$ is the combined volume and time average,
and denote by $Nu(t)$ the instantaneous Nusselt number obtained by averaging only over the 
volume, not time. In addition, we measure the small-scale Reynolds number $Re$, based on 
the horizontal length scale and the root-mean-square vertical velocity, by taking the time 
average of the instantaneous Reynolds number $Re(t)=\sqrt{\langle w^2 \rangle _{\rm vol}}$
(see Sec.~\ref{sec:methods}). Both $Nu$ and $Re$ are commonly used to quantify turbulent 
convection since their statistics are known to converge quickly in the nonlinear regime 
following the initial exponential growth phase of the convective instability, see 
e.g. \citet{maffei2021inverse}. This remains true even in the presence of an inverse 
cascade leading to a slow growth in the horizontal kinetic energy. Figure~\ref{fig:Nu_Re_Ro}$(a)$ 
displays the time series of the instantaneous Nusselt number at $\ekman=10^{-10}$ for three different 
reduced Rayleigh numbers $\widetilde{Ra}=40,60,80$, showing increasing turbulent heat flux (mean 
and fluctuations) with increasing $\widetilde{Ra}$. Rescaling the Nusselt numbers by the well-known 
turbulent scaling law \citep{kJ12} for rapidly rotating convection $Nu-1\sim\widetilde{Ra}^{3/2}$ 
(see panel $b$), the data collapse satisfactorily for $\widetilde{Ra}=60,80$, while the least 
turbulent simulation at $\widetilde{Ra}=40$ shows a small mismatch. The corresponding 
theoretical scaling prediction for the Reynolds number is $Re\sim \widetilde{Ra}$, which we 
use to rescale our Reynolds number data. 

Averaging over time series like those shown in the top row of figure~\ref{fig:Nu_Re_Ro}, we 
obtain the Nusselt and Reynolds number statistics in panels $c$ and $d$ of 
figure~\ref{fig:Nu_Re_Ro}, shown versus the Taylor number $Ta =Ek^{-2}$. Specifically, four 
sets of simulations at constant $\widetilde{Ra}=40,60,80,120$ are shown for a wide range 
of $10^2\leq Ta\leq 10^{30}$. Triangles in different orientations for each $\widetilde{Ra}$ 
show the observed average $Nu$ and $Re$, while error bars indicate the measured standard 
deviation about those mean values. Horizontal dashed lines indicate, for each $\widetilde{Ra}$, 
the values corresponding to the reduced NHQG equations, with a shaded region indicating the 
corresponding standard deviation measured in the NHQG equations. At small $Ta$ (large $Ek$), for 
the given $\widetilde{Ra}$, the flow is only weakly supercritical and therefore features only 
small $Nu$ and $Re$. A break is seen in the curves around $Ta=10^6$ ($\ekman=10^{-3}$), where the 
bulk convective Rossby number equals one, marked by faint vertical dashed lines, indicating that 
the bulk flow begins to be rotationally constrained. The inverse cascade sets in around $Ta=10^{8}$ 
for $\widetilde{Ra} =40$, and closer to $Ta = 10^{10}$ for larger $\widetilde{Ra}=60,80,120$ 
(hollow circular markers). This is in agreement with the literature, see e.g. 
\citet{favier2014inverse}. As $Ta$ increases ($\ekman$ decreases) further from this point, 
an overshoot in $Nu$ is observed leading to a maximum in the convective heat transport 
for the given control parameter $\widetilde{Ra}$, whose amplitude and location depend 
on $\widetilde{Ra}$. This is followed by convergence within one standard deviation to the
value obtained in the NHQG equations at yet larger $Ta$, beyond a threshold indicated by a 
second set of vertical dashed lines, one for each $\widetilde{Ra}$, corresponding to convective 
Rossby number in the boundary layer of order unity, a condition equivalent to $\ekman={\sigma}^{3/2} 
\widetilde{Ra}^{-15/4}$  \citep{julien2012heat}, i.e., the rotation rate where the boundary 
layer also becomes rotationally constrained. These regime boundaries are equivalent to those 
shown in the top panel of figure~\ref{fig:overview_parameter_space}. A similar structure 
is observed in the Reynolds number data, which collapse slightly less satisfactorily under 
the rescaling (this is known to be a consequence of the presence of the large-scale vortex 
\citep{maffei2021inverse,oliver2023small}). At small $Ta$, the flow is only weakly 
supercritical and therefore the Reynolds number is small. It increases with $Ta$ up to a 
maximum, beyond which it decreases to converge (within one standard deviation) to the 
value obtained from the NHQG equations. Since Taylor numbers in most laboratory experiments 
or unrescaled DNS have been limited to $Ta\lesssim 10^{14}-10^{16}$, this $Nu$ and $Re$ vs $Ta$ 
dependence has not previously been described. In particular, the presence of a maximum in 
the heat transport and turbulence intensity at a finite Taylor (or equivalently, Ekman) 
number as well as the statistical convergence to the NHQG equations are reported here for 
the first time. 

In addition to the \textit{a priori} (bulk/boundary layer) convective Rossby numbers shown 
by the dashed vertical lines in panels $c$ and $d$ of figure~\ref{fig:Nu_Re_Ro}, 
panels $e$ and $f$ show the \textit{a posteriori} Rossby number $Ro$ computed as the 
root-mean-square of the ratio $\left\|\boldsymbol{u \cdot \nabla u}_\perp\right\|
/(2\Omega \left\|\boldsymbol{u}_\perp\right\|)$ in the bulk and in the boundary layer, 
where it takes distinct values (see Appendix~\ref{sec:app_C}). When shown versus 
$\varepsilon=\ekman^{1/3}$, the standard small parameter of rotating convection, the data,
while scattered, clearly show that the bulk Rossby number increases monotonically with $\ekman$ as 
expected, crossing unity near the observed onset of the inverse cascade. It is interesting that 
this occurs at Ekman numbers which are two orders of magnitude smaller than what may naively be 
expected based on the convective Rossby number $Ro_{conv}=1$. When the same data is plotted 
against the rescaled parameter $Ek^{1/3} \widetilde{Ra}^{5/4}$, it collapses to a 
remarkable degree, both in the bulk and in the boundary layer. We stress that this is 
somewhat surprising, since this effective small parameter is obtained explicitly from 
the derivation of the reduced equations for the boundary layer \citep{julien2012heat,kJ12}. 
However, we note that the bulk flow develops cyclone--anticyclone symmetry at similar Ekman 
numbers as the boundary layer, which is another empirical link between the bulk and boundary 
layer flows observed here. The collapse of the data in figure~\ref{fig:Nu_Re_Ro}$e$,$f$ implies 
in particular that when $Ek^{1/3} \widetilde{Ra}^{5/4} = 1$, below which the average Nusselt 
and Reynolds numbers converge to the NHQG data, there is a well-defined associated critical 
Rossby number in the boundary layer, which is found to be approximately $0.2$. This is where 
the boundary layer loses rotational constraint and becomes unstable to strong ageostrophic 
motions with pronounced vertical gradients. These additional vertical gradients are associated 
with additional dissipation contributions. In steady state, the dissipation rates of kinetic 
energy and temperature variance are directly linked to the Nusselt number, via the so-called 
power integrals. In the RRRiNSE formulation, these read
\begin{subequations}
\label{eq:power_ints}
\begin{align}
\langle |\nabla_\perp \theta|^2 \rangle 
	+ \varepsilon^2 \langle (\partial_z \theta)^2 \rangle 
	+ \langle (\partial_z {\Theta})^2\rangle 
	=&  Nu -1, \label{eq:power_int_T}\\ 
	\langle |\partial_x \boldsymbol{u}|^2 + |\partial_y \boldsymbol{u}|^2\rangle 
	+\varepsilon^2\langle |\partial_z \boldsymbol{u}|^2\rangle 
	=&  \frac{\widetilde{Ra}}{\sigma^2} (Nu-1)\label{eq:power_int_kin},
\end{align}
\end{subequations}
where $\langle \cdot \rangle$ is the combined time and volume average over the whole domain. 
Directly measuring all contributions in our simulations, we find that both balances are 
approximately satisfied (indicating that nonstationarity is weak, see Appendix~\ref{sec:app_C}), 
with important contributions to the dissipation of temperature variance originating in the 
boundary layers near $Ek^{1/3} \widetilde{Ra}^{5/4} = 1$. The increase in the Nusselt number 
away from the small $\ekman$ NHQG limit can therefore be understood as a consequence of increased 
dissipation due to enhanced (vertical and horizontal) gradients, at least in part due to the 
strong anticyclonic vortical structures appearing in the boundary layer at $Ek=10^{-8}$ 
(which is near $Ek^{1/3} \widetilde{Ra}^{5/4} = 1$ for $\widetilde{Ra}=120$), 
cf.~figure~\ref{fig:flow_visualisation_combined}.
\subsection{Boundary layer structure}
\begin{figure}
    \centering
\begin{overpic}
	[width=0.4925\textwidth]{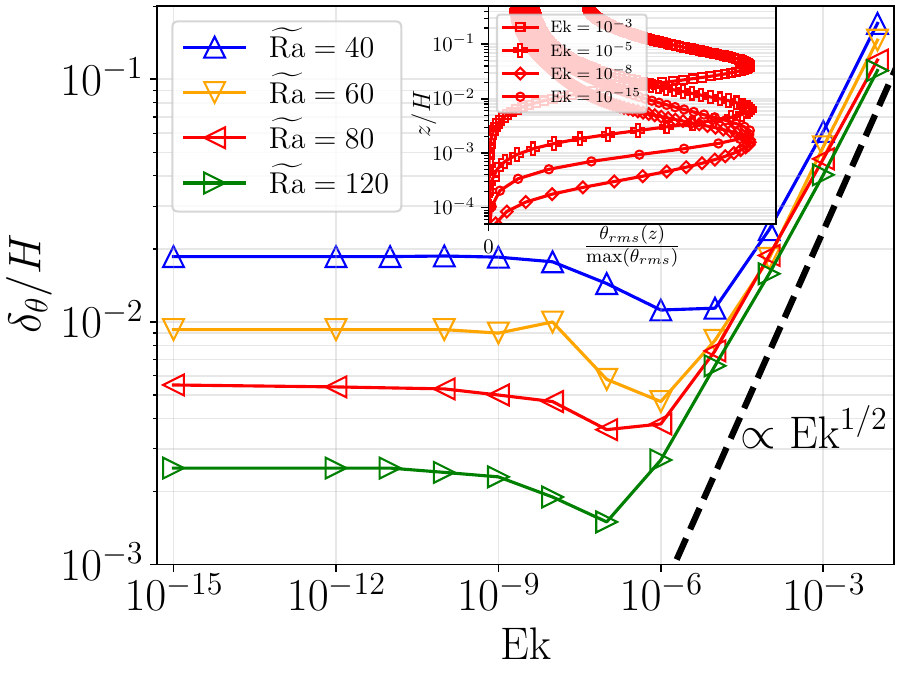} \put (0,70) {$(a)$}
\end{overpic}
\begin{overpic}
	[width=0.4925\textwidth]{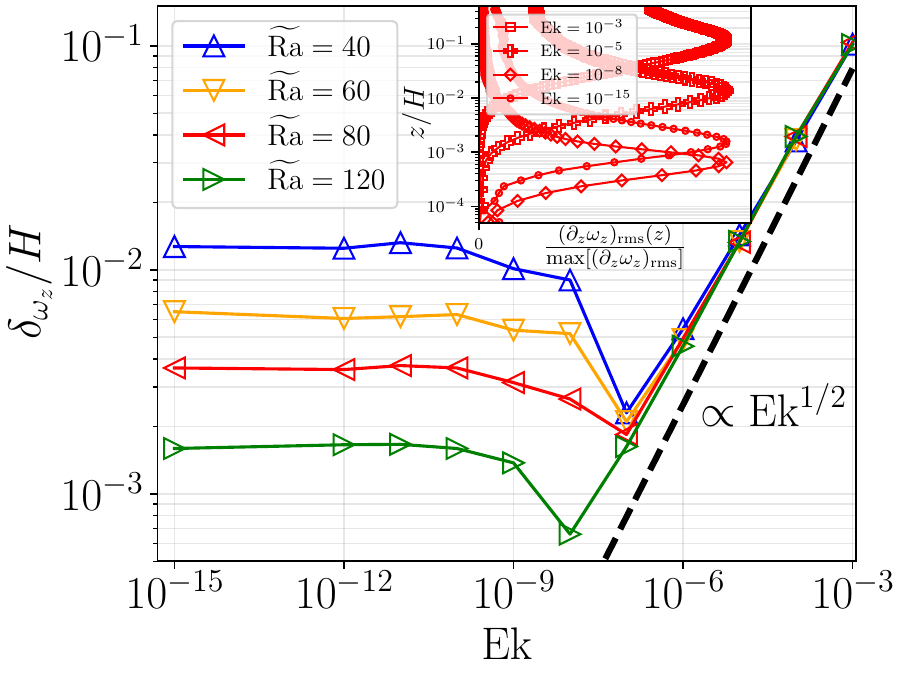}\put (0,70) {$(b)$}
\end{overpic}
\\   
\begin{overpic}
	[width=0.49\textwidth]{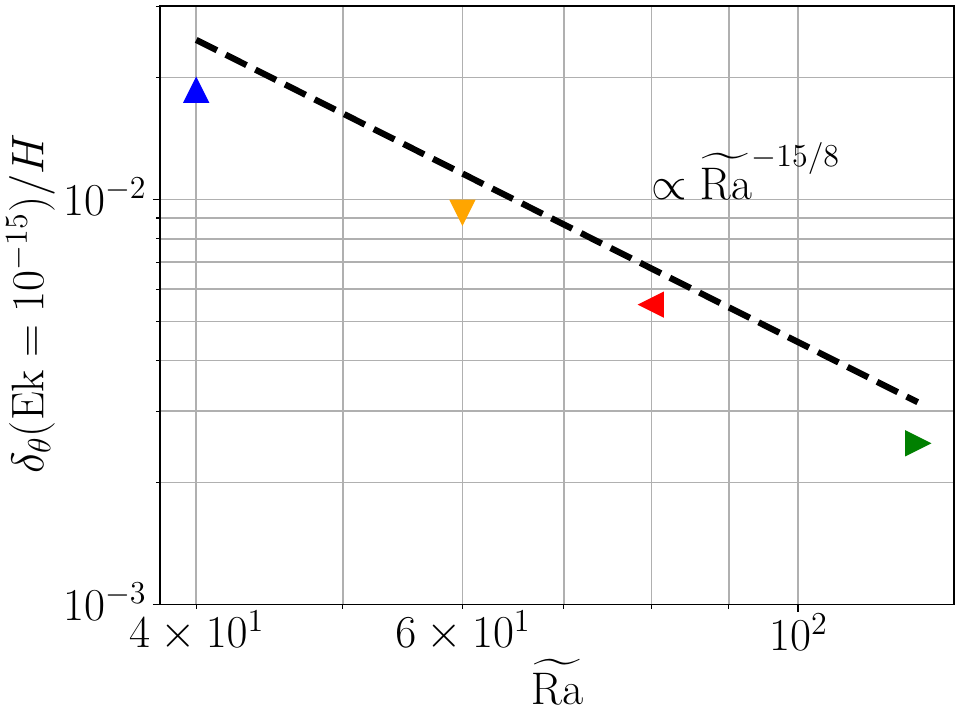} \put (0,70) {$(c)$}
\end{overpic}
    \caption{\raggedright 
Boundary layer structure for a range of $Ek$ and $\widetilde{Ra}$. Panel $a$: 
log--log plot of the non-dimensional widths of the thermal boundary layer $\delta_\theta/H$, 
given by the location in $z$ of the maxima in the vertical root-mean-squre (r.m.s.) profiles
of $\theta$ shown in the inset, versus $\ekman$ for different $\widetilde{Ra}$. Dashed line 
indicates a $Ek^{1/2}$ power law. Panel $b$: log--log plot of the momentum boundary layer 
thickness $\delta_{\omega_z}/H$ based on the location of the maxima of the r.m.s. profile
of $\partial_z \omega_z$ shown in the inset. Panel $c$: log--log plot of the non-dimensional 
thermal boundary layer thickness $\delta_\theta/H$ at $Ek=10^{-15}$ compared with the power 
law $\widetilde{Ra}^{-15/8}$ predicted by the NHQG equations \citep{kJ12}. 
    }
\label{fig:width_of_thermal_and_momentum_bls}
\end{figure}

From the above analysis, it is apparent that the boundary layers play a nontrivial role 
in the transition from small to large $\ekman$, controlling the departure from the NHQG branch. 
To further characterize the boundary layer structure, we measure and analyze vertical 
root-mean-square profiles of different quantities as a function of $z$. 
Figure~\ref{fig:width_of_thermal_and_momentum_bls}$(a)$ shows a log--log plot of the thermal
boundary layer depth $\delta_\theta$ computed as the location of the maximum in the 
root-mean-square vertical profile of $\theta$ (shown in the inset for different $\ekman$ at 
$\widetilde{Ra}=120$). At small Ekman numbers, $\delta_\theta$ converges to a constant 
value for a given reduced Rayleigh number, corresponding to the NHQG limit. Panel $c$ confirms 
that the thermal boundary layer depth scales with $\widetilde{Ra}^{-15/8}$, which is 
the scaling obtained from reduced NHQG equations for the boundary layer \citep{julien2012heat,kJ12}.
As $\ekman$ is increased close to the point where the boundary layer loses rotational support 
($\ekman = \sigma^{3/2}\widetilde{Ra}^{-15/4}$, not shown), the boundary layer depth undershoots before 
increasing with a power law not far from $Ek^{1/2}$ (and steeper than $\ekman^{1/3}$, not shown). 
Figure~\ref{fig:width_of_thermal_and_momentum_bls}$(b)$ shows an alternative measure of 
the boundary layer thickness, based on the root-mean-square vertical profile of 
$\partial_z \omega_z$. The inset shows sample profiles at $\widetilde{Ra}=120$ for different 
Ekman numbers. A local maximum is seen in these profiles, which we use to define a 
momentum boundary layer thickness $\delta_{\omega_z}$. Since we are considering stress-free 
boundaries, where no linear Ekman layer is present, the existence of this momentum boundary 
layer is nontrivial. We interpret $\delta_{\omega_z}$ as a measure of how far the effect of 
the stress-free boundary penetrates into the bulk of the flow. 

We have furthermore verified that the momentum boundary layer seen here in the RRRiNSE 
formulation is also present in the reduced NHQG equations (not shown), although to our 
knowledge this has not been described previously. Like $\delta_\theta$, at sufficiently 
small $\ekman$, $\delta_{\omega_z}$ becomes constant, while at larger values of $\ekman$, there 
is again an undershoot followed by a power-law scaling regime where $\delta_{\omega_z}$ 
increases approximately as $\ekman^{1/2}$. The approximate $\ekman^{1/2}$ scaling, which is 
observed both in terms of $\delta_\theta$ and in $\delta_{\omega_z}$ and is also observed 
for no-slip boundary conditions \citep{song2024direct}, is surprising in the present case, 
given that we consider stress-free boundaries, which do not lead to the formation of linear 
Ekman layers. While a boundary layer thickness proportional to $\ekman^{1/2}$ is theoretically 
predicted for the transitional thermal boundary layer \citep{kJ12}, the presence of such a 
scaling at significantly larger Ekman numbers, where the boundary layer flow is no longer 
rotationally dominated, goes beyond this prediction and the precise origin of this scaling 
law remains to be elucidated.
 
\subsection{Alternative cuts through parameter space}
\begin{figure}
    \centering
    \begin{overpic}
        [width=0.49\textwidth]{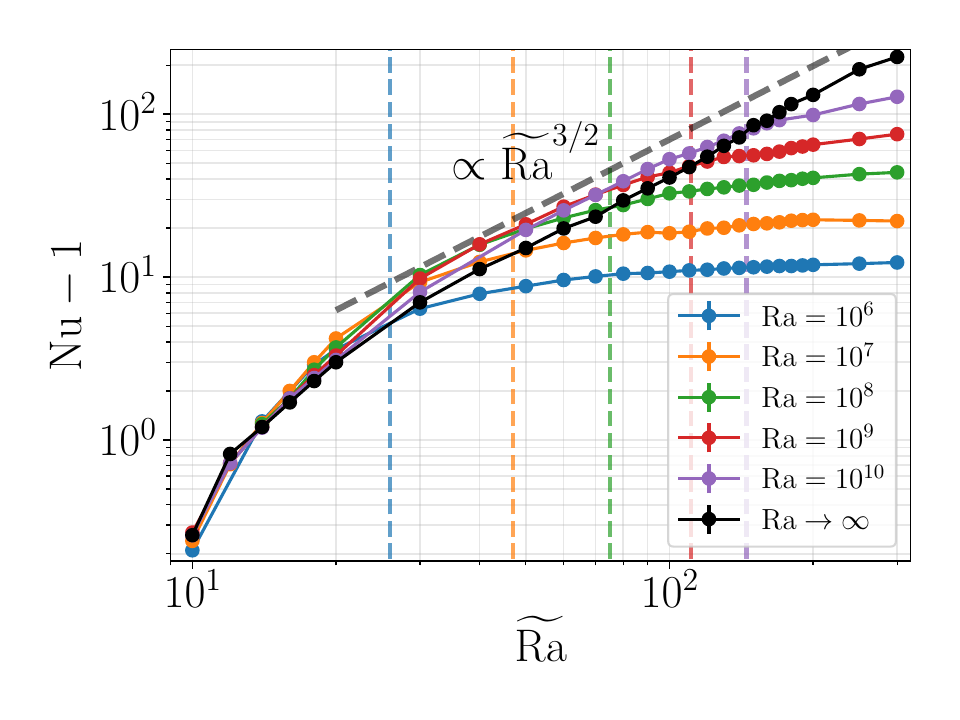}
	\put (5,65) {$(a)$}
    \end{overpic}
    \begin{overpic}
        [width=0.49\textwidth]{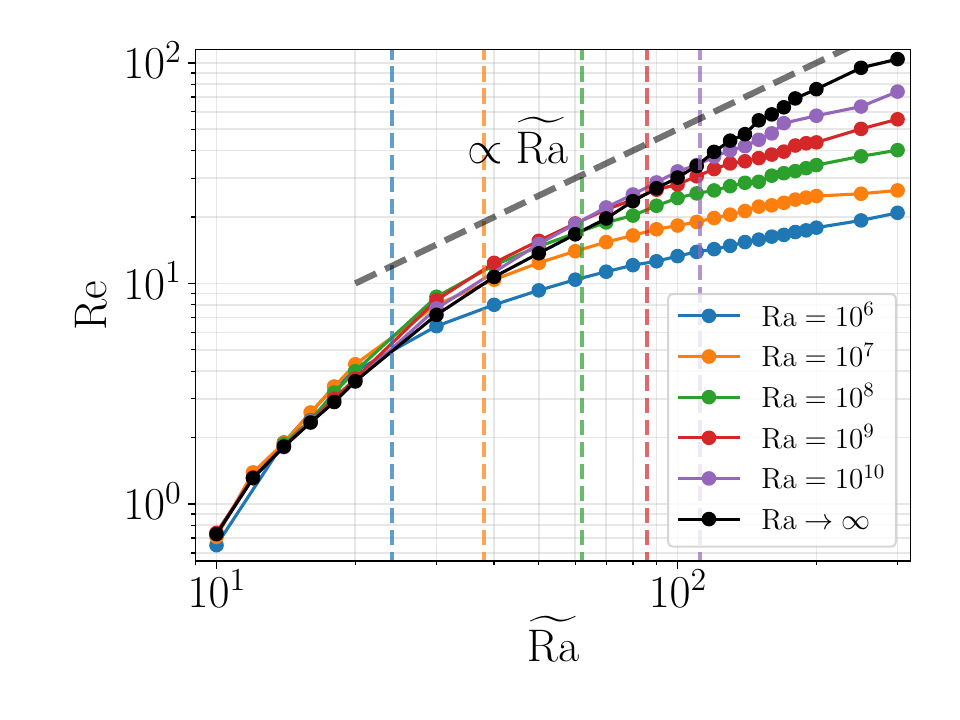}
	\put (5,65) {$(b)$}
    \end{overpic}
    \begin{overpic}
        [width=0.49\textwidth,height=0.36\textwidth]{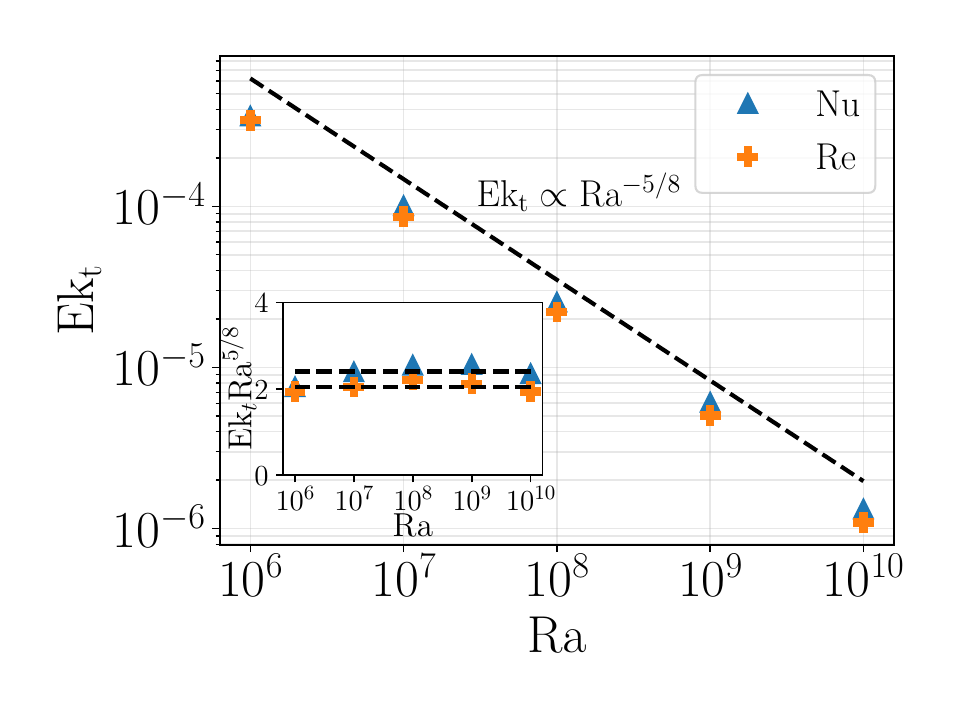} \put 
	    (5,65) {$(c)$}
    \end{overpic}
  \begin{overpic}
	  [width=0.49\textwidth]{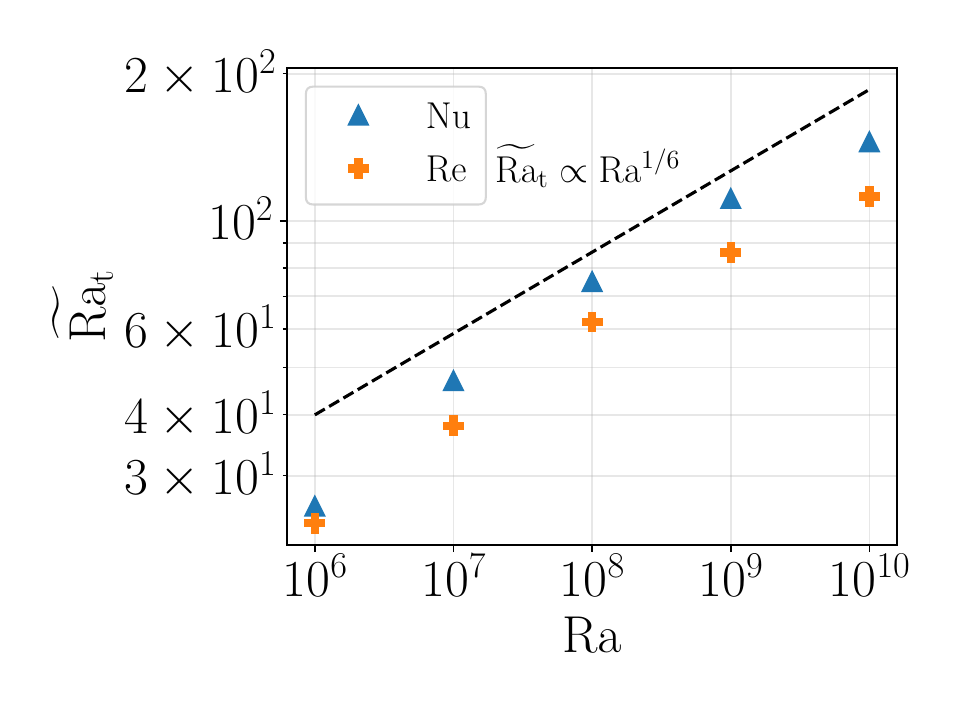} \put (5,65) {$(d)$}
  \end{overpic}
    \begin{overpic}
        [width=0.49\textwidth]{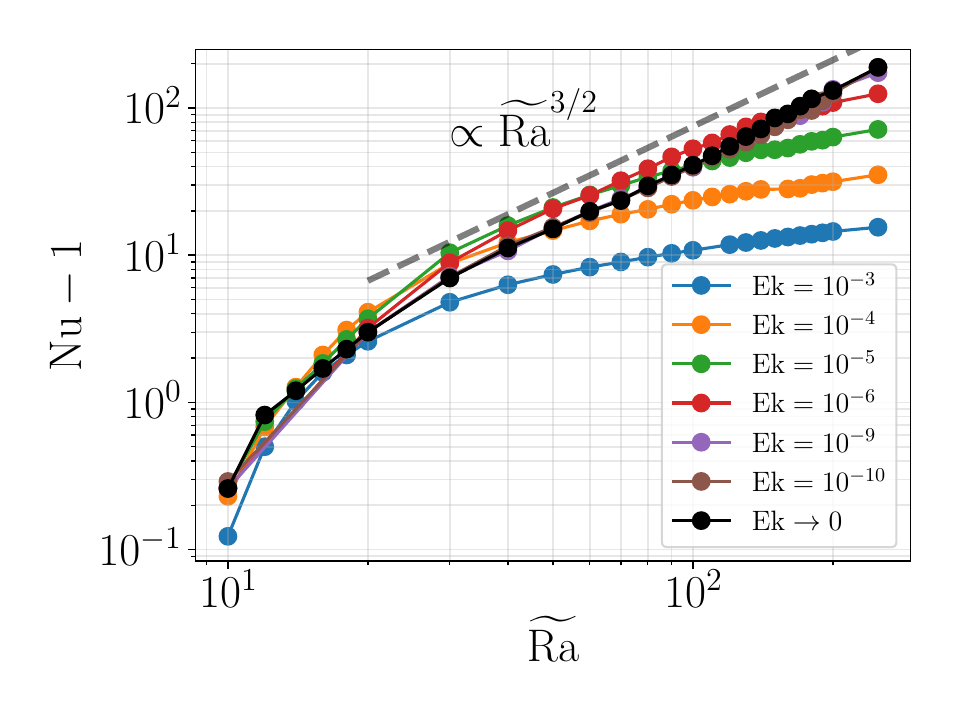}
	\put (5,65) {$(e)$}
    \end{overpic}
    \begin{overpic}
       [width=0.49\textwidth]{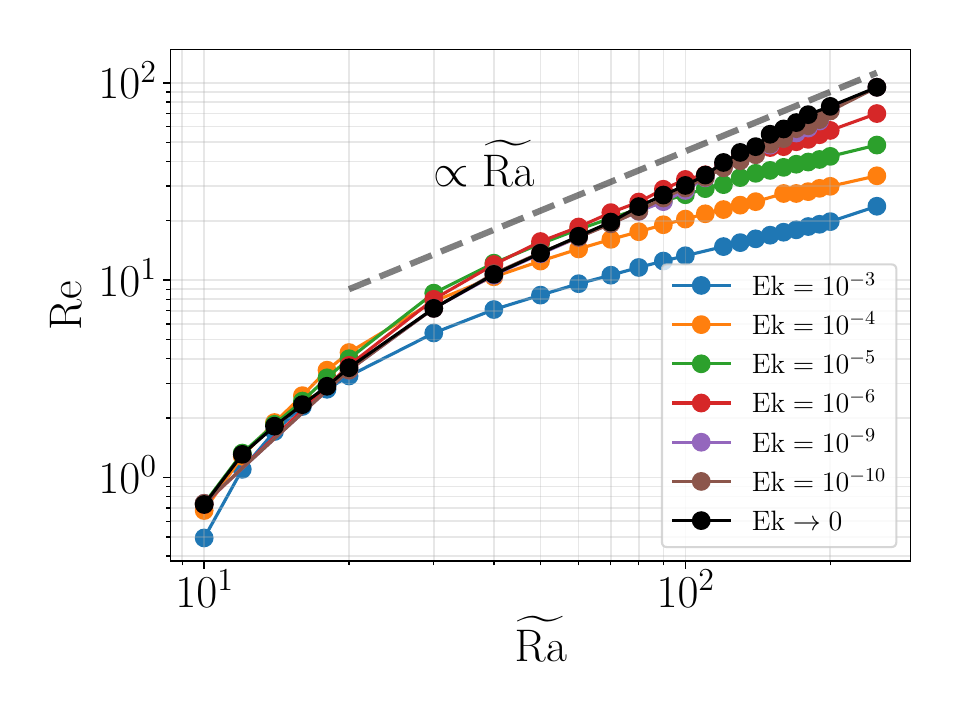} 
       \put (5,65) {$(f)$}
    \end{overpic}
    \caption{\raggedright
Nusselt number (panel $a$) and Reynolds number (panel $b$) versus the reduced 
Rayleigh number $\widetilde{Ra}$ at fixed bare Rayleigh number $Ra$. Dashed lines indicate 
intersection of a given $Ra=const.$ set with the asymptotic limit. Transitional Ekman number 
(panel $c$) and reduced Rayleigh number (panel $d$) versus the bare Rayleigh 
number $Ra$. $Nu-1$ (panel $e$) and $Re$ (panel $f$) versus $\widetilde{Ra}$ at fixed Ekman number.
\label{fig:Nu_Re_vs_rRa_Ra_fixed}}
\end{figure}

The results presented thus far were obtained along a particular cut through the physical 
control parameter space, namely varying $Ek$ at fixed $\widetilde{Ra}$. This is the natural 
approach in the framework of the RRRiNSE. By contrast, keeping $\widetilde{Ra}$ fixed in 
an experimental setup or in an unrescaled DNS requires tuning two parameters. In such contexts, 
it is instead more natural to vary $Ra$ at fixed $Ek$ (equivalent to varying $\widetilde{Ra}$ 
at fixed $Ek$) or vary $Ek$ at fixed $Ra$ (equivalent to varying $\widetilde{Ra}$ at fixed $Ra$). 
To make contact with existing experimental and numerical studies, the corresponding cuts through 
parameter space are shown in figure~\ref{fig:Nu_Re_vs_rRa_Ra_fixed}. Panel $a$ shows $Nu -1$ 
versus $\widetilde{Ra}$ for different sets of RRRiNSE simulations, each with $Ra$ fixed to 
a different value between $Ra=10^6$ and $Ra=10^{10}$, as well as the corresponding values 
from the NHQG equations ($Ra \to \infty$) at the same $\widetilde{Ra}$. The NHQG data 
approximately follow the turbulent $\widetilde{Ra}^{3/2}$ scaling law, cf. 
figure~\ref{fig:Nu_Re_Ro}, while at finite $Ra$, there is a well-defined transitional 
value $\widetilde{Ra}_t$ of $\widetilde{Ra}$ where $Nu -1$ intersects the $Ra\to \infty$ limit, 
and subsequently follows a shallower slope. This structure of the Nusselt number flattening 
out beyond a threshold $\widetilde{Ra}=\widetilde{Ra}_t$ is robust among the datasets of runs 
with different $Ra$. Qualitatively similar behavior was recently reported for RRRBC with 
no-slip boundaries \citep{song2024direct}, although the eventual low $\ekman$ regime remains 
elusive in that case. Panel $b$ shows the corresponding Reynolds number data for the 
same simulations. As discussed earlier, the corresponding turbulent scaling law 
$Re\propto \widetilde{Ra}$ has proved to be more elusive owing to the impact of the 
inverse cascade, see \citet{maffei2021inverse}, \citet{oliver2023small} and 
figure~\ref{fig:Nu_Re_Ro}. Panel $b$ shows a similar structure to panel $a$, in that 
the $Re$ curves intersect the curve corresponding to the NHQG equations at a 
(numerically slightly different) threshold value for each $\widetilde{Ra}$. 
Since these runs are performed at constant $Ra$, each $\widetilde{Ra}$ is equivalent to 
an Ekman number $Ek$. Therefore $Ra_t$ is equivalent to a transitional Ekman 
number $Ek_t$. 

Panel $c$ shows $Ek_t$, corresponding to the vertical dashed lines in panels $a$ and $b$, 
indicating that the data are quantitatively compatible with the scaling law $Ek_t\propto Ra^{-5/8}$.
This scaling law corresponds to the theoretical prediction \citep{julien2012heat} for the 
Ekman number at which the boundary layer loses rotational support. Here, we recover this 
scaling from a simple measurement of the Nusselt and Reynolds numbers. 
An equivalent representation of the same data is shown in panel $d$ in terms of the reduced 
Rayleigh number $\widetilde{Ra}$ versus the bare Rayleigh number $Ra$, which also reveals 
satisfactory agreement with the corresponding theoretical scaling prediction 
$\widetilde{Ra}_t \propto Ra^{1/6}$. 

Panels $e$ and $f$ of figure~\ref{fig:Nu_Re_vs_rRa_Ra_fixed} show a third possible cut 
through the parameter space of rotating convection (qualitatively resembling results presented 
in \citet{song2024direct}), holding $Ek$ fixed and changing $Ra$ (or equivalently changing 
$\widetilde{Ra}$). The turbulent scaling laws with $\widetilde{Ra}$ compare similarly well 
with these datasets and exhibit similar changes in scaling in the Nusselt and Reynolds 
numbers with $\widetilde{Ra}$ at finite $Ek$, up to $\ekman=10^{-6}$, while for $\ekman=10^{-9}$ and 
$\ekman=10^{-10}$ the values of $\widetilde{Ra}$ attainable in our simulations were not 
sufficiently high to observe a departure from the curve corresponding to the NHQG equations.

The flattening of the Nusselt number curves with increasing $\widetilde{Ra}$ at finite Ekman 
numbers is a robust signal that could potentially be reproduced in laboratory experiments. 
Our results in figure~\ref{fig:Nu_Re_vs_rRa_Ra_fixed} qualitatively resemble similar findings 
recently obtained for RRRBC with no-slip boundaries \citep{song2024direct}.

\section{Discussion}
\label{sec:disc}
The results described above show that the RRRiNSE formulation, based on low Ekman number 
asymptotics, allows efficient direct numerical simulations (DNS) far beyond the current state 
of art in laboratory experiments or unrescaled DNS, down to $\ekman=10^{-15}$ and below, which is 
comparable to the estimated values of $\ekman$ in the outer core of the Earth and the convection 
zone of the Sun. Specifically, this indicates that the RRRiNSE enable DNS at Ekman numbers 
which are over six orders of magnitude smaller than the smallest value obtained previously, 
in an impressive effort and at great computational cost, by means of unrescaled DNS 
\citep{song2024direct,Song_Shishkina_Zhu_2024}. Using the RRRiNSE formulation, we revealed 
rich new physics in this previously inaccessible parameter regime. First, we uncovered a 
transition at $\ekman \approx \sigma^{3/2}\widetilde{Ra}^{-15/4}$ ($Ek\approx 10^{-9}$ for $\widetilde{Ra}=120$) 
from a regime at larger $\ekman$ where the depth-averaged flow features a strong large-scale 
cyclone and a weak diffuse anticyclone towards a regime characterised by cyclone--anticyclone 
symmetry and a large-scale vortex dipole, in agreement with the predictions of the NHQG equations. 
Second, we identified a nontrivial transition in the boundary layer dynamics corresponding to 
the loss of rotational support in the boundary layer (confirming a previously untestable 
theoretical prediction of \citet{kJ12}), a transition accompanied by the emergence of strong, 
albeit short-lived, anticyclonic structures near the $\ekman$ threshold that weaken 
as $\ekman$ increases, leading to dominance of strong cyclonic structures. It is interesting 
to note that the cyclone--anticyclone symmetry-breaking in the bulk approximately coincides 
in $Ek$ with the boundary layer transition, although a theoretical explanation of this
observation remains unavailable. While flows in the Earth's atmosphere are characterised 
by only moderately small Rossby numbers and are therefore dominated by strong cyclones, our 
results suggest that in the interior of the Earth or other celestial bodies, cyclones and 
anticyclones may be statistically of equal strength. 

We showed quantitatively that the time-averaged Nusselt and Reynolds numbers in steady 
state reflect these regime transitions, taking values consistent with the NHQG limit 
for $\ekman \lesssim \sigma^{3/2}\widetilde{Ra}^{-15/4}$, while overshooting as $\ekman$ is increased at 
fixed $\widetilde{Ra}$, reaching maximum heat transport near the value of $\ekman$ where the 
bulk Rossby number reaches unity, and decreasing to close to zero as $\ekman$ increases further 
at fixed $\widetilde{Ra}$ due to reduced supercriticality. The increase in the Nusselt number 
was shown to be tied to increased dissipation associated with a change in the boundary layer 
flow morphology. The boundary depth was quantified in terms of the thermal fluctuations and 
vertical vorticity, revealing that, at $\ekman \lesssim \sigma^{3/2}\widetilde{Ra}^{-15/4}$, the boundary layer 
depth becomes $\ekman$-independent, taking the value associated with the NHQG equations, but 
undershooting for $\ekman \gtrsim \sigma^{3/2}\widetilde{Ra}^{-15/4}$ before increasing approximately with
an Ekman layer-like scaling $Ek^{1/2}$ as $\ekman$ increases. Owing to the presence of stress-free 
boundary conditions, this behavior cannot be explained by a linear Ekman layer and is therefore 
a nonlinear effect. Finally, we have considered alternative cuts through the parameter space, 
with one set of runs varying $\ekman$ at fixed $\rayleigh$ and another set of runs varying $\rayleigh$ 
at fixed $\ekman$. This procedure revealed that at finite Ekman numbers, the Nusselt and Reynolds 
numbers remain close to the values found in the NHQG equations as $\widetilde{Ra}$ increases, 
but start to deviate from them at a value of $\widetilde{Ra}$ that depends on $Ra$ in a 
way consistent with boundary layer theory \citep{julien2012heat}. We mention that 
the NHQG equations apply to systems with no-slip boundaries since in the limit $\ekman\to0$ 
the no-slip boundaries become effectively stress-free \citep{kJ12}. At finite $\ekman$, 
however, Ekman boundary layers and the associated Ekman pumping necessarily modify the 
boundary layer structure, an effect that can be included in the NHQG equations 
following \citet{kJ16}.

The results presented here are based on the RRRiNSE reformulation of the equations and 
explain why previous state-of-the art DNS and laboratory experiments could not reach the 
parameter regime for observing the transition to fully rotationally constrained dynamics. 
The RRRiNSE formulation therefore opens the door to the further exploration of the
parameter regime of very small but finite Ekman and Rossby numbers, highly relevant to planetary, 
satellite and stellar interiors. This is all the more timely since current and future 
observational missions prompted by the last Planetary Decadal Survey~\citep{planetDecadalSurv}, 
including the recently launched ESA's JUICE (JUpiter Icy Moons Explorer) 
mission \citep{grasset2013jupiter} and NASA's Europa Clipper mission~\citep{clipper_NASA_2020} 
will provide new data requiring interpretation based on faithful model simulations at parameters 
as close as possible to realistic values. Our results are a first step in this direction.

The RRRBC as studied here, with antiparallel gravity and rotation axis is an appropriate, 
albeit idealised, model of the North Pole regions of the celestial bodies listed in 
table~\ref{Table:Paramo}. Owing to the latitudinal non-uniformity of rotating convection 
in spherical shells \citep{gastine2023latitudinal}, the RRRiNSE formulation cannot be 
straightforwardly generalised to that setting, but an investigation of the RRRiNSE with 
misaligned gravity and rotation axes could provide important insights in this regard. 
While RRRiNSE enables simulations at realistic values of $\ekman$, many challenges remain: 
the increasing resolution constraint for increasing levels of turbulence cannot be 
circumvented, and for simplicity our simulations considered stress-free boundaries as 
well as Prandtl number $\sigma=1$ while ignoring other important ingredients such as magnetic 
fields. To further test the applicability of our findings to planets, icy moons and stars, 
future work must also investigate how the regime transitions described in this study 
are influenced by additional effects, including fluids with Prandtl numbers different 
from one, compressibility, internal heating~\citep{barkerApJ14, bouillautPNAS21, hadjerciJFM24}, 
and strong magnetic fields, as well as alternative choices of boundary conditions. 
The doors to these explorations are now open thanks to the RRRiNSE formulation. 
Comparison of the RRRiNSE numerical results with future laboratory experiments would 
also be highly desirable.

\backsection[Acknowledgments] 
{
K. J. passed away before this manuscript was finalised. We have attempted to present 
the results of our collaboration in accordance with his high standards. Any errors or 
misinterpretations remain our own. Part of the writing of this manuscript was done while 
one of the authors (AvK) was a Staff Member at the Woods Hole Oceanographic Institution 
Geophysical Fluid Dynamics summer program. This research used the Savio computational cluster 
resource provided by the Berkeley Research Computing program at the University of California 
Berkeley (supported by the UC Berkeley Chancellor, Vice Chancellor for Research, and 
Chief Information Officer). This research also utilised the Alpine high performance 
computing resource at the University of Colorado Boulder. Alpine is jointly funded by the 
University of Colorado Boulder, the University of Colorado Anschutz, and Colorado State 
University. Data storage for this project was supported by the University of Colorado 
Boulder PetaLibrary. This project was also granted access to computational 
resources of TGCC under the allocation 2024-A0162A10803 made by GENCI, and to resources 
of PMCS2I (P\^ole de Mod\'elisation et de Calcul en Sciences de l’Ing\'enieur et de l’Information) 
of Ecole Centrale de Lyon. We thank Daria Holdenried-Chernoff for pointing out useful
references on dynamo simulations at low Ekman number and Rudie Kunnen for helpful comments. 
This research was funded, in whole or in part, by Agence Nationale de la Recherche 
(Grant ANR-23-CE30-0016-01). 
A CC-BY public copyright license has been applied by the authors to the present document and 
will be applied to all subsequent versions up to the Author Accepted Manuscript arising 
from this submission, in accordance with the grant’s open access conditions.
}
\backsection[Funding]{This work was supported by the National Science Foundation 
(Grants DMS-2009319 and DMS-2308338 (KJ), Grants DMS-2009563 and DMS-2308337 (EK)), 
by the German Research Foundation (DFG Projektnummer: 522026592) and by Agence 
Nationale de la Recherche (Grant ANR-23-CE30-0016-01).}

\backsection[Supplementary movies] {Supplementary movies are available at \href{https://doi.org/10.1017/jfm.2025.290}{https://doi.org/10.1017/jfm.2025.290}.}

\backsection[Declaration of interests] {The authors report no conflict of interest.}

\backsection[Author ORCIDs]{\\
A. van Kan, https://orcid.org/0000-0002-1217-3609;\\
K. Julien, https://orcid.org/0000-0002-4409-7022;\\
B. Miquel, https://orcid.org/0000-0001-6283-0382;\\ 
E. Knobloch, https://orcid.org/0000-0002-1567-9314.}

\appendix

\begin{figure}
\centering
\hspace{1cm} $(a)$ \hspace{7cm} $(b)$\\
 \includegraphics[width=\textwidth]{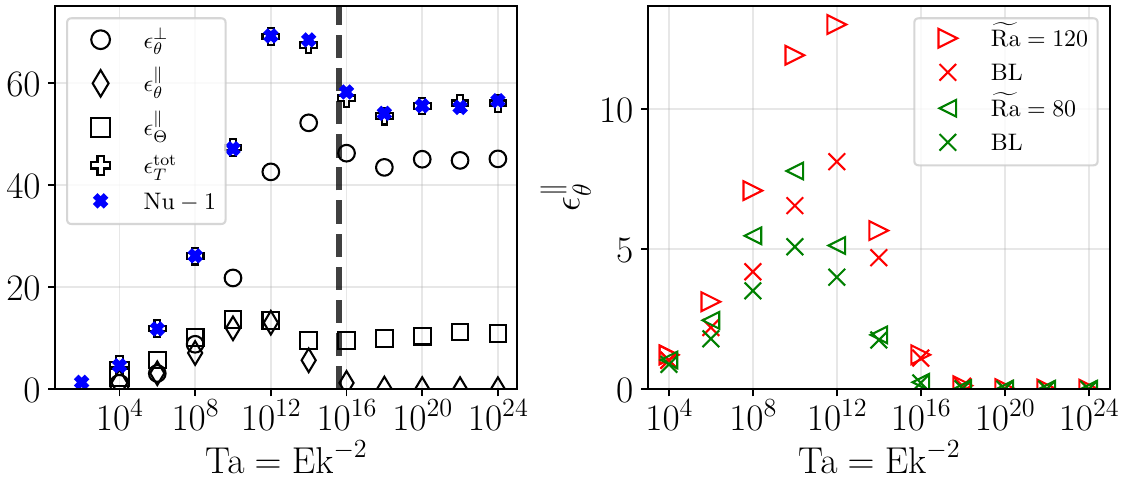}
 \hspace{1cm} $(c)$ \hspace{7cm} $(d)$\\
\includegraphics[width=\textwidth]{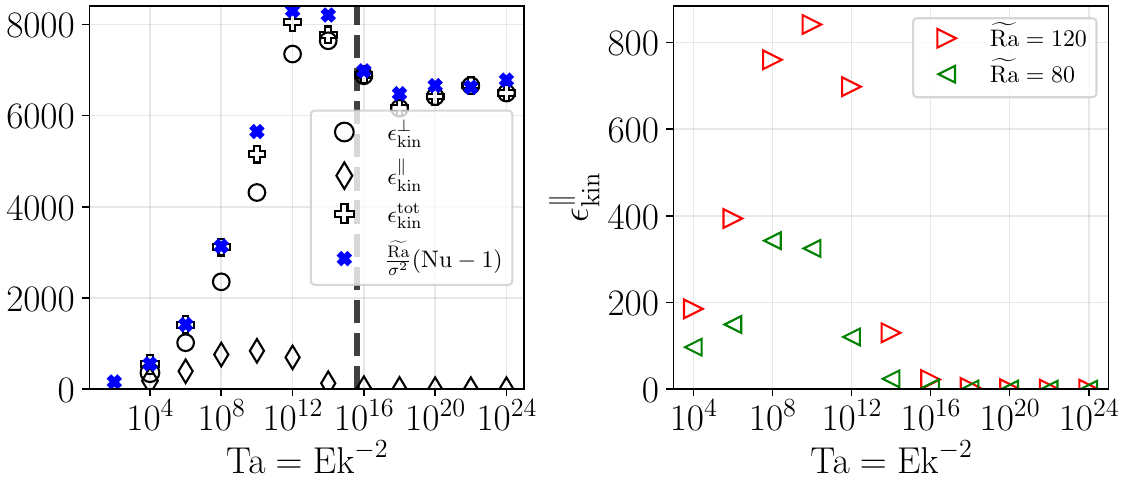}
 \caption{\raggedright
Panel $a$ shows the dissipation budget of the temperature variance arising from 
vertical flow variations at $\widetilde{Ra}=120$ versus $Ta=Ek^{-2}$, 
with $\epsilon_\theta^\perp \equiv \langle |\nabla_\perp \theta|^2\rangle$, 
$\epsilon^\parallel_\theta \equiv \epsilon^2 \langle (\partial_z \theta)^2 \rangle$ 
and $\epsilon_\Theta^\parallel\equiv \langle (\partial_z \Theta)^2 \rangle$, showing that the 
power-integral equation~(\ref{eq:power_int_T}) of the main text is well satisfied and hence 
that a statistically steady state has been reached. The overshoot in $Nu-1$ is primarily 
due to increased $\epsilon_\theta^\perp$ and $\epsilon_\theta^\parallel$. 
Panel $b$ shows the dissipation due to vertical variations in $\theta$ at $\widetilde{Ra}=80,120$ versus 
$Ta=Ek^{-2}$ together with the boundary layer contributions (marked by {\large $\times$}, 
integrated over a depth of $2\delta_\theta$). 
Panel $c$ shows the kinetic energy budget 
showing that the contributions from the horizontal 
($\epsilon_{kin}^\perp \equiv \langle | \nabla_\perp \boldsymbol{u} |^2 \rangle$) 
and vertical gradients 
($\epsilon_{kin}^\parallel\equiv\varepsilon^2\langle |\partial_z \mathbf{u})|^2\rangle$) 
approximately add up to $(Nu-1) \widetilde{Ra}/\sigma^2$, as predicted by (\ref{eq:power_int_kin}) 
of the main text. A small mismatch is seen at $Ek \lesssim 10^{-5}$ since the emerging 
large-scale vortex has not fully saturated in amplitude. Panel $d$ shows the contribution 
from vertical variation to the kinetic energy dissipation versus $Ta$ at $\widetilde{Ra}=80,120$.
\label{fig:dissipation}}
\end{figure}
\section{Additional details on flow statistics and structure\label{sec:app_C}}
Supplementary Movies 1 to 3 show the evolution of the vertical vorticity $\omega_z$, the 
temperature perturbation $\theta$ and the vertical velocity $w$, respectively, at the top 
of the momentum boundary layer $z=\delta_{\omega_z}$ for $Ek=10^{-15}$ and $\widetilde{Ra}=120$. 
Supplementary Movies 4 to 6 provide the same information for $Ek =10^{-8}$. At $Ek=10^{-15}$, 
all fields display close to zero skewness and cyclones are of approximately equal strength and 
structure as anticyclones. In contrast, at $Ek=10^{-8}$, the boundary layer is characterised 
by the presence of strong, albeit short-lived, anticyclonic vortical structures typically with 
a shielded structure that impose a clear signature on $\theta$ and $w$, associated with cold 
spots in the centre where the fluid descends while rising within a surrounding ring.
Figure~\ref{fig:dissipation} shows an analysis of the dissipation budget predicted by the power-integral 
relations given in (\ref{eq:power_ints}) in the main text, as a function of the 
Taylor number $Ta\equiv Ek^{-2}$. Panel $a$ shows that in the statistically stationary state 
at $\widetilde{Ra}=120$ the sum of the dissipation contributions from the temperature 
variations due to gradients in the horizontal and vertical directions sums up approximately
to $Nu-1$ as predicted by (\ref{eq:power_int_T}). The departure from the approximately constant 
Nusselt number at small $Ek$ is seen to be associated with an increase in dissipation due to 
both horizontal and vertical gradients. Panel $b$ shows the contributions from vertical gradients, 
which primarily stem from the boundary layers near the departure from the NHQG limit. Panel $c$ 
shows the kinetic energy dissipation budget as a function of the Taylor number at 
$\widetilde{Ra}=120$. The sum of the dissipation of kinetic energy due to vertical 
and horizontal gradients shows a satisfactory overall agreement with $(Nu-1)\widetilde{Ra}/\sigma^2$,
as predicted by (\ref{eq:power_int_kin}). However, while for $Ta \lesssim 10^{8}$, where no 
inverse energy cascade is observed, the agreement is close to perfect, it is notable that for 
$Ta \gtrsim 10^{10}$, the measured kinetic energy is slightly below the value predicted by 
(\ref{eq:power_int_kin}). This is because the large-scale vortex is still slowly growing in 
amplitude via the condensation process. Panel $d$ shows the contributions from vertical 
gradients, which show a similarly drastic transition from negligibly small to finite values, 
contributing to the observed overshoot in the Nusselt number.

\begin{figure}
        \includegraphics[height=0.38\linewidth]{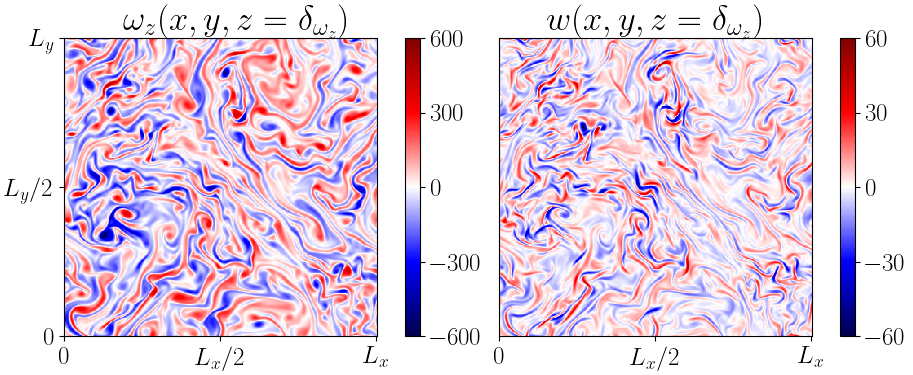}
        \includegraphics[height=0.38\linewidth]{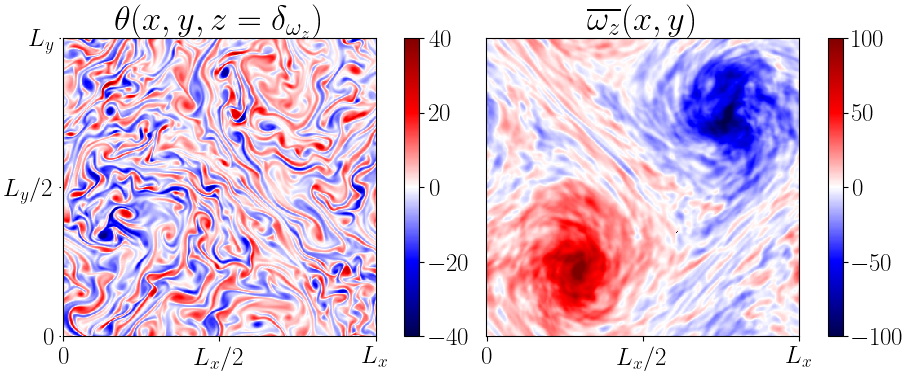}
    \caption{\raggedright
Snapshots of the flow state obtained via direct numerical simulation of 
the RRRiNSE at $Ek=10^{-24}$ and $\widetilde{Ra}=120$ in terms of $\omega_z$, $w$, $\theta$ at 
the top of the boundary layer $z=\delta_{\omega_z}$, and the depth-averaged vorticity 
field $\overline{\omega_z}$ revealing a symmetric large-scale vortex dipole.}
\label{fig:flow_field_Ek1em24}
\end{figure}
The RRRiNSE formulation remains numerically stable at even lower Ekman numbers than those 
discussed in the main text.  Figure~\ref{fig:flow_field_Ek1em24} illustrates this in terms 
of the flow field obtained for $\widetilde{Ra}=120$ at $Ek=10^{-24}$. The flow morphology is 
indistinguishable from that at $Ek=10^{-15}$. Specifically, in the boundary layer, there is 
no cyclone--anticyclone asymmetry and the $\omega_z$ and $\theta$ fields are nearly perfectly 
correlated. Moreover, the barotropic vertical vorticity displays a counter-rotating vortex 
dipole at the domain scale, similar to what is seen in two-dimensional turbulence. 
While the observed flow structure and the associated flow statistics at $Ek=10^{-24}$ are 
close to $Ek=10^{-15}$ for the case considered here, more analysis is needed to determine 
the lowest Ekman numbers attainable with RRRiNSE given finite machine precision, and to 
characterize the breakdown of the method.

\begin{figure}
    \centering
    \includegraphics[width=0.5\linewidth]{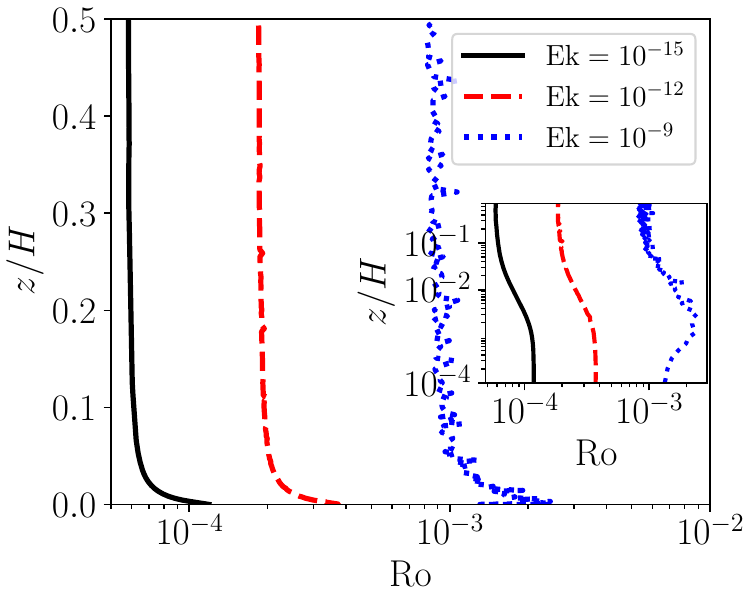}
    \caption{\raggedright
Vertical profiles of the r.m.s. local Rossby number
$|\boldsymbol{u \cdot \nabla u}|/(2\Omega |\boldsymbol{u}_\perp|)$ versus non-dimensional 
height at $\widetilde{Ra}=120$ for Ekman numbers $Ek = 10^{-15},10^{-12},10^{-9}$. Inset 
shows same data in log--log representation, showing that bulk and boundary layer have distinct
values of the Rossby number. At $Ek=10^{-9}$, there is a local maximum in the Rossby number 
observed near the top of the thermal boundary layer. The values shown in the main text 
are obtained by averaging over the thermal boundary layer volume.}
    \label{fig:rms_Ro_profiles}
\end{figure}
At a given Ekman number, the importance of the Coriolis force compared to nonlinear 
acceleration is not uniform in the vertical direction. To illustrate this, 
figure~\ref{fig:rms_Ro_profiles} shows the vertical r.m.s. profiles of the Rossby number, 
computed based on the ratio $\left|\boldsymbol{u \cdot \nabla u}_\perp\right|/
(2\Omega |\mathbf{u}_\perp|)$. This quantity has a well-defined value in the bulk of the flow, 
but assumes a distinct value in the boundary layer, as can be seen in the doubly logarithmic 
representation in the inset. These two distinct values characterising the bulk flow and 
the boundary layer are shown in panels $e$ and $f$ of figure~\ref{fig:Nu_Re_Ro} in the main text.

\bibliography{bridging_low_rossby}
\bibliographystyle{jfm}

\end{document}